\documentclass[useAMS,usenatbib]{mn2e}
\usepackage{times}
\usepackage{graphicx}

\title[Detecting ``Temperate'' Jupiters]{Detecting ``Temperate'' Jupiters:  The Prospects of Searching for Transiting Gas Giants in Habitability Zones}
\author[S.W. Fleming, S.R. Kane, P.R. McCullough, F.R. Chromey] {S.W. Fleming$^1$\thanks{scfleming@astro.ufl.edu}, S.R. Kane$^{1,2}$, P.R. McCullough$^3$, F.R. Chromey$^4$\\
$^1$Department of Astronomy, University of Florida,  211 Bryant Space Science Center, Gainesville, FL 32611-2055, USA\\
$^2$Michelson Science Center, Caltech, MS 100-22, 770 South Wilson Avenue, Pasadena, CA 91125, USA\\
$^3$Space Telescope Science Institute, 3700 San Martin Drive, Baltimore, MD, 21218, USA\\
$^4$Department of Physics and Astronomy, Vassar College,  124 Raymond Ave, Poughkeepsie, NY, 12604, USA}
\begin{document}


\pagerange{\pageref{firstpage}--\pageref{lastpage}} \pubyear{2008}

\maketitle

\label{firstpage}

\begin{abstract}
Wide-field photometric surveys in search of transiting extrasolar planets are now numerous and have met with some success in finding Hot Jupiters.  These transiting planets have very short periods and very small semi-major axes, facilitating their discovery in such surveys.  Transiting planets with longer periods present more of a challenge, since they transit their parent stars less frequently.  This paper investigates the effects of observing windows on detecting transiting planets by calculating the fraction of planets with a given period that have zero, one (single), two (double), or $\ge$3 (multiple) transits occurring while observations are being taken.  We also investigate the effects of collaboration by performing the same calculations with combined observing times from two wide-field transit survey groups.  For a representative field of the 2004 observing season, both XO and SuperWASP experienced an increase in single and double transit events by up to 20-40$\%$ for planets with periods $14 < P < 150$ days when collaborating by sharing data.  For the XO Project using its data alone, between 20-40$\%$ of planets with periods 14-150 days should have been observed at least once.  For the SuperWASP Project, 50-90$\%$ of planets with periods between 14-150 days should have been observed at least once.  If XO and SuperWASP combined their observations, 50-100$\%$ of planets with periods less than 20 days should be observed three or more times.  We find that in general wide-field transit surveys have selected appropriate observing strategies to observe a significant fraction of transiting giant planets with semimajor axes larger than the Hot Jupiter regime.  The actual number of intermediate-period transiting planets that are detected depends upon their true semimajor axis distribution and the signal-to-noise of the data.  We therefore conclude that the investment of resources needed to investigate more sophisticated photometry calibrations or examine single and double transit events from wide-field surveys might be a worthwhile endeavour.  The collaboration of different transit surveys by combining photometric data can greatly increase the number of transits observed for all semimajor axes.  In addition, the increased number of data points can improve the signal-to-noise of binned data, increasing the chances of detecting transiting extrasolar planets.
\end{abstract}


\section{Introduction}
\label{introduction}
Many groups have searched for transiting extrasolar planets, e.g., VULCAN \citep{bor2001}, STARE \citep{bro1999}, OGLE \citep{uda2002}, SuperWASP \citep{pol2006}, HAT \citep{bak2004}, XO \citep{mcc2005} and TrES, \citep{alo2004}, and have been
successful at detecting so-called Hot Jupiters
\citep{uda2002,tor2004,bou2004,kon2004,alo2004,mcc2006,bak2007a,cam2007}, etc.  These planets are
characterised by extremely short periods and small semi-major axes, increasing their photometric detectability.  
Planets orbiting in the Habitability Zone (HZ) of main-sequence stars have periods significantly longer than the orbital periods of Hot Jupiters and lower geometric transit probabilities.  Ground-based, wide-field
photometric surveys typically examine a region of sky for only
a few months, and therefore the maximum number of HZ planet transits observable per year is only 1 or 2.  These transits are not sufficient to
unambiguously measure the actual period.  However, combined with radial velocity follow-up
and further photometry measurements, it might be possible to use these transits as the starting point for detecting transiting ``Temperate'' Jupiters.

Because follow-up telescope time is competitive and expensive, only those targets with many detected transits and very well-defined periods are usually considered for spectroscopic follow-up.  In this paper, we define a ``single transit event'' as a transiting companion for which only one transit was observed or detected, likewise, a ``double transit event'' is one where only two transits were observed or detected.  Some of the single or double transit events could be intermediate-period transiting companions, or they could be short-period companions where additional transits were not detected.  In this paper, we make a distinction between an ``observation'' and a ``detection''.  An ``observation'' means a telescope was observing a star while a transit was occurring, while a ``detection'' means that in addition, the signal-to-noise of the data is sufficient for data reduction software to detect the event.  Detecting a transiting planet depends on details like the photon noise of the data, the number of transits observed, the severity of red noise and choice of algorithm to correct it and quality of the photometric calibration.  For example, \emph{observations} can be made during poor weather, while \emph{detections} depend sensitively on weather conditions.  This paper therefore investigates the prospects of \emph{observing} an intermediate-period planet (defined in this paper as 14 days $< P < $ 730 days), an essential first step in \emph{detecting} such planets via transit surveys.  Our definition of an intermediate-period planet encompasses the HZ for most F-M spectral-type, main-sequence stars.  Detection probability is the most important quantity in real wide-field transit search operations, since it helps determine the yield and the speed that the discoveries will be made.

The probability of not observing a given transiting planet is determined solely by the times of observations.  Given a set of dates for which observations were taken, it is possible to estimate as a function of period the probability of observing a transiting planet exactly once, exactly twice, at least three times, or missing it entirely, by effectively integrating over all possible values of orbital phase.  Assuming each measurement has equivalent signal-to-noise and that all measurements are internally consistent (i.e. measured relative to the same reference) then even measurements without a transit event can be useful.  Given a sufficient number of measurements, enough period/phase combinations can be ruled out to yield a set of discrete solutions for the period and phase.  Combined with the expected number of transits observed using the numerical technique presented here, it is possible to estimate how likely the single or double transit event might be an intermediate-period companion and hence worthy of further follow-up.  The results of this analysis are particularly relevant to single or double-transit events with extremely high signal-to-noise ratios, such as eclipsing binary stars for ground-based telescopes or gas giant planets for space-based missions like \emph{Kepler} and \emph{CoRoT}, where the signal-to-noise is sufficient to allow very precise modelling of a single transit to be performed, minimising potentially expensive follow-up time and resources.

Section \ref{motivation} briefly discusses some of the scientific motivation for detecting and studying HZ gas/ice giants.  Section \ref{phase1} estimates the transit probabilities for intermediate-period planets and describes how an upper limit on the period can be estimated for single and double transit events.  Section \ref{code} describes the numerical technique used while Sections \ref{indi} and \ref{sharingsection} present the results for the projects analysed individually and collectively.  Section \ref{assump} addresses some of the assumptions used in our technique, and Section \ref{summary} summarises and concludes.

\section{Scientific Motivation}
\label{motivation}
Planets with intermediate-length periods offer insights to several questions regarding extrasolar planetary science.  Data from the many radial velocity surveys in progress already shows there is an observed lack of gas and ice giants with periods between 14 days and a few years \citep{but2006}.  The lack of planets at these orbital distances offers an important testbed for planetary migration theory.  The cause of these planets halting migration before reaching the Hot Jupiter regime and their relative frequency is a question that can be addressed through a partnership of theory and observation.

Extrasolar gas/ice giants with these intermediate-length periods have the potential to exist in the Habitability Zone of their parent star (which we refer to as ``Temperate Jupiters'' in this paper).  One of the benefits of transiting extrasolar planets is that it is possible to use precision timing of the transit to search for other planets, satellites, or ring systems \citep[e.g.,][]{sar1999,doy2002}.  Due to the close proximity of Hot Jupiters to their parent stars, large natural satellites are not dynamically stable over significant periods of time \citep{bar2002}. However, if Temperate Jupiters have systems of natural satellites then they could represent potentially habitable worlds outside our solar system, in addition to HZ planets.

A necessary first step in studying HZ natural satellites is to observe how many HZ gas/ice giants exist and what their properties are like.  While radial velocity surveys have already detected extrasolar planets potentially in their parent star's HZ, transiting HZ planets offer the capability of directly measuring the planet's radius, albedo and chemical composition.  Depending on the natural satellite's orbital parameters, it is possible to measure its radius and mass, as well as estimate its structural composition.  It could be that some of the first terrestrial worlds discovered in the HZ of a main-sequence star are found as natural satellites of Temperate Jupiters rather than as HZ planets.  Indeed, transiting HZ natural satellites of gas and ice giants might be more easily detectable and more readily studied in the near future because the parent planets act as signposts for when and where to search.  \citet{rob2007}, for example, have discovered two gas giant planets orbiting with semi-major axes between 1-2 AU in nearly-circular orbits, and are examples of stars that might have habitable natural satellites.

The detection of a transiting Temperate Jupiter would offer interesting studies in its own right.  It is possible to measure a transiting planet's orbital inclination relative to our line-of-sight.  When combined with \emph{m sin(i)} planet mass measurements from radial velocity, the extrasolar planet's mean density can be derived.  Already this has been done for several extrasolar planets with some surprising results, including planets with larger-than-expected radii and gas giants with very massive solid cores.  Fig. 3 of \citet{bak2007b} offers an example of the wide range of transiting extrasolar planet mean densities.  However, all known transiting planets are Hot Jupiters that reside in extreme environments with very strong radiation fields.  Detecting a transiting extrasolar planet with these ``intermediate periods'' would allow tests of planetary atmosphere models without examining only extrasolar planets in strongly irradiated environments.  It is also possible to detect the chemical composition of transiting planet atmospheres via high-resolution differential spectroscopy \citep{cha2002,vid2003,vid2004}, producing many interesting results applicable to planetary atmosphere models and interior structure theories.  Hot Jupiters feature unique effects due to their extreme temperatures, including atmospheric escape \citep[e.g.,][]{vid2003, lia2003}.  The detection of a transiting Temperate Jupiter would allow for the first chemical composition measurements of a non-extreme atmosphere outside our own system.

Wide-field photometric surveys will detect many other interesting objects besides extrasolar planets, including many types of variable stars.  Diluted eclipsing binaries in particular can be difficult to distinguish from transiting planets.  Grazing or blended eclipsing binaries offer a chance at testing the detection capability of a transit survey, since their transit depths can be similar to that of transiting extrasolar planets.  In addition, the number of low-mass detached eclipsing binaries is not numerous, despite their use in determining the stellar mass-radius relationship \citep{lop2004}, the importance of which affects nearly all fields of astrophysics, including extrasolar planet searches.  They are also excellent tests for stellar evolution models \citep{las2002} and as nearby extragalactic distance indicators \citep{wyi2001}.  The radii of known eclipsing low-mass binaries and brown dwarfs have generally been larger than models predict, e.g. \citet{bay2006}, and in the case of a brown dwarf - brown dwarf binary, the smaller brown dwarf was found to have (somewhat surprisingly at the time) a greater effective temperature \citep{sta2006}, meaning that more systems are needed to derive better models of the atmospheres of low-mass stars and brown dwarfs.

\section{Initial Detection}
\label{phase1}
\subsection{Probabilities of a Transit}
\label{subsecprobs}
The probability of observing a transiting planet is greater for Hot
Jupiters than those with longer periods since they transit their parent stars more frequently and have a greater geometric probability of being favourably
inclined for observation.  Following \citet{bor1984}, the geometric probability
of an inclination leading to a transit is given by
\begin{equation}
P_t = R_* / a
\label{prob_tran}
\end{equation}
where $R_*$ is the radius of the star, $a$ is the planet's orbital radius, and the orbit is assumed to be circular.  This is only the
probability of the planet transiting and not of actually detecting the
transit.  Other factors such as the signal-to-noise of the data determine
the detectability of a transit.  Jovian planets in the HZ of a star
will therefore have a lower geometric probability of transiting their
parent star, making them an intrinsically rarer find.

The determination of the exact boundaries of the HZ is a very complex
matter involving climate theories, stellar evolution, atmospheric
models and planetary orbital parameters.  For simplicity, in this work we assume that the present-day HZ of a star is such that the star's bolometric irradiance at the surface of the planet is equal to that of the Sun's at the Earth, i.e., the HZ is centred around a distance given by
\begin{equation}
a_\mathit{HZ} = 1.~\mathrm{AU} \sqrt{\frac{L}{L_\odot}}
\label{HZdist}
\end{equation}
where $L$ is the bolometric luminosity of the star and $L_\odot$ is
the solar bolometric luminosity.  More detailed studies of the Habitable Zone can be found in \citet{kas1993}, \citet{tur2003}, \citet{jon2006}, and references therein.  \citet{goufis2003} used the above
assumption to show that currently astrometry offers the best means of
detecting HZ planets orbiting massive stars, primarily A and F spectral types.  Astrometric detection signals depend directly on the size of the
planetary orbit, and the HZ for more massive stars
is located at a greater distance.  Complementary to that, the method of transit photometry is more
sensitive to planets with small semi-major axes since 
the probability of a planet transiting is inversely proportional to
its semi-major axis.  The transit depth is also larger for stars with smaller radii.
Therefore, transit photometry is best able to detect HZ planets around less massive stars, especially G,K,M dwarfs.  In fact,
\citet{goupep2003} have shown that although current transit searches
are most sensitive overall to planets orbiting G-type stars, they are
best able to detect HZ planets around M-type stars.

The geometric transit probability and the approximate semi-major axis of a planet in the HZ for various spectral types are
calculated using Equations \ref{prob_tran} and \ref{HZdist}, respectively, and shown in Table
\ref{probtable}.  It also contains the period of a Jupiter-mass planet in a circular orbit via
Kepler's Third Law and the stellar parameters found in the table.  As mentioned in Section \ref{assump}, the assumption of circular orbits means these geometric probabilities are approximate lower limits.  The values presented in Table \ref{starprop} are only representative but are sufficient for demonstrating geometric transit probabilities of HZ planets.
\begin{table}
\begin{minipage}{84mm}
\caption{\sevensize{Geometric probabilities ($P_t$) of transits and orbital periods (P) for planets in the HZ of select spectral type stars.  $M_*$ is the primary star's mass, $L$ is the star's bolometric luminosity, $R_*$ is the star's radius and $a_{HZ}$ is the distance to the centre of the Habitable Zone for that particular star.}\label{probtable}}
\begin{tabular}{@{}ccccccc}
\hline
 Spectral Type & $M_*^{~a}$\footnotetext[1]{\citet{all2000}} & $\log{\frac{L}{L_{\odot}}}^{~b}$ \footnotetext[2]{\citet{cox2000table}}& $R_*^{~a}$&
 $a_\mathit{HZ}$ & $P_t$ & $P$
\\
     &               [$M_{\odot}$] &    & [$R_{\odot}$] & [AU] & [\%]
 & [years]
\\
\hline
O5 &  60  & 5.7 & 12 & 710   &  0.0080 &   2400\\
B0 &  17.5 & 4.3  & 7.4  & 140   &  0.025 & 400   \\
B5 &  5.9 & 2.9    & 3.9  & 28.    &  0.065 & 61   \\
A0 &  2.9  & 1.9     & 2.4  & 8.9   &  0.13 & 16    \\
A5 &  2.0  & 1.3     & 1.7  & 4.5   &  0.18 & 6.7    \\
F0 &  1.6  & 0.8    & 1.5  & 2.5   &  0.28 & 3.1    \\
F5 &  1.4  & 0.4    & 1.3  & 1.6   &  0.38 & 1.7    \\
G0 &  1.05  & 0.1    & 1.1  & 1.1   &  0.46 & 1.1    \\
G5 &  0.92  & -0.1   & 0.92  & 0.89  &  0.48 & 0.87    \\
K0 &  0.79  & -0.4   & 0.85  & 0.63  &  0.63 & 0.56    \\
K5 &  0.67  & -0.8   & 0.72  & 0.40  &  0.84 & 0.31    \\
M0 &  0.51  & -1.2  & 0.60  & 0.25  &  1.1 & 0.17     \\
M2 &  0.40  & -1.5  & 0.50  & 0.18  &  1.3 & 0.12     \\
M5 &  0.21  & -2.1 & 0.27  & 0.090 &  1.4 & 0.059    \\
M8 &  0.06  & -3.1 & 0.10  & 0.028 &  1.7 & 0.019    \\
\hline
\label{starprop}
\end{tabular}
\end{minipage}
\end{table}

For most spectral types, the probability of detecting a transiting HZ planet via a single-site transit survey is small because the orbital periods are greater than the duration of a typical observing interval for a field (nominally $\sim$90 days).  Even if such a survey observes continuously the planet may never transit during the three-month observing interval.  Collaboration amongst projects or multi-site observatories at different longitudes can expand a typical observing interval from 2-3 months for a given field out to 6+ months, increasing the chances of observing an intermediate-period transiting planet.  This will be discussed more in Section \ref{sharingsection}.  For the purposes of this paper, I explore period ranges from 1 to 730 days (this includes the HZ range for stars that are the focus of radial velocity and transit surveys, i.e., mid-F through M spectral types).  ``Intermediate-period'' generally refers to periods ranging from 14-730 days in this paper, unless stated otherwise.

\subsection{Information From the Light Curve}
Efforts are underway to detect the transits of intermediate-period
planets already known by radial velocity measurements \citep{sea2003}.  Indeed, several planets have been shown not to transit\footnote{See {\tt
    http://www.ucolick.org/$\sim$laugh/} for an up-to-date list.}.  HD 17156b with a period of 21.2 days is an example of a planet that was found by \citet{barb2007} to transit after it was discovered via radial velocity \citep{fis2007}.  
Another approach is to perform wide-field searches photometrically and confirm candidates with radial velocity follow-up.  A Jupiter-sized planet around a G2 V star produces a transit depth of approximately 1\%, which is detectable using small-aperture telescopes.

Most transit detecting algorithms, including the BLS
method \citep{kov2002}, could detect single transit events that might be intermediate-period companions.  However, since there is no periodicity shown and the light curve is ill-defined, the period cannot be
tightly constrained based solely on the light curve.  Once such a detection is made there are a few properties
that can be approximated.  First,
the spectral type of the parent star needs to be determined and the stellar mass estimated, for example, by obtaining medium-resolution spectroscopy of the target star.  Once the
spectral type of the parent star is known, the radius can be estimated via the stellar mass-radius relationship $R_* = kM_*^x$ where $k=1$ and
$x=0.8$ for F-K main-sequence stars \citep{cox2000}.  Providing the spectral
resolution is high enough, it is possible to distinguish diluted,
double-lined eclipsing binaries at this point.  Finding low-mass,
detached eclipsing binaries has several applications, as discussed in
Section \ref{motivation}.

The depth of the transit can be approximated by taking an average of the
transit data points.  Once the depth is approximated, it is possible
to calculate the radius of the planet using the relationship $
R_p = R_* \sqrt{d}$ where $R_p$ is the radius of the planet, and $d$ is the depth of the transit.

Assuming a circular orbit for simplicity, limits on the transit duration can be
obtained by measuring the time between the measurements before and after the transit has occurred.  This constraint will
depend upon the observing window of the data, i.e., on the
number and length of gaps in the photometry.  Section \ref{assump} discusses the effects of eccentricity on the transit duration and the effect it has on our results.  Using Kepler's Third
Law, the equation for the transit duration ($D_t$), and the constraint
on $D_t$, it is possible to solve the system of two equations and
obtain an initial value for the period, $P$ \citep[see][for details on
  analytic solutions to transit equations]{seag2003}.  The transit
duration is given by
\begin{equation}
D_t = \frac{P}{\pi} \arcsin{\left(\frac{R_*}{a} \sqrt{\frac{(1+R_p/R_*)^2-((a/R_*) \cos i)^2}{1-\cos^2 i}}\right)}
\label{tteq}
\end{equation}
where $i$ is the orbital inclination to our line-of-sight and $a$ is the orbital semimajor
axis.  Kepler's Third Law for a circular orbit is given by
\begin{equation}
P^2 = \frac{4 \pi^2 a^3}{G(M_*+M_p)}
\label{k3leq}
\end{equation}
where $G$ is the Gravitational Constant and $M_*$ and $M_p$ are the masses of the star and planet,
respectively. Solving for $a$ in Equation \ref{k3leq} and
substituting into Equation \ref{tteq} with $i=90^\circ$, it is
possible to obtain a relationship between $D_t$ and $P$:
\begin{equation}
D_t = \frac{P}{\pi} \arcsin{\left(\frac{(R_*+R_p)(4\pi^2)^{1/3}}{(G(M_*+M_p)P^2)^{1/3}}\right)}
\label{guesspeq}
\end{equation}
Assuming $M_p << M_*$, it is possible to evaluate Equation
\ref{guesspeq} for $P$ using the upper-limit estimate on $D_t$.  The ability to constrain the orbital period is dependent upon the signal-to-noise of the data, the observing window and the error on the estimates of the other parameters.  For analysing initial light curves of targets from wide-field surveys, such analytical analysis can be difficult to apply due to the noise and gaps inherent in the data.  However, placing an upper limit on the period can be important when considering follow-up options for potential intermediate-period planet candidates.

\begin{figure}
\includegraphics[width=84mm]{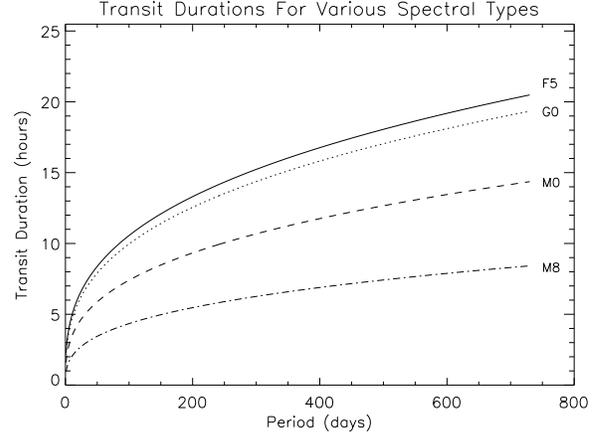}
\caption{Transit duration vs. period for a Jupiter-mass planet in a circular orbit around various spectral type stars.}
\label{durations}
\end{figure}

Fig. \ref{durations} shows the transit duration for a Jupiter-mass planet in a circular orbit around an F5, G0, M0 and M8 spectral type star as a function of orbital period.  The masses and radii of the stars are taken from Table \ref{starprop} and Eqn. \ref{durationmaster} is used to estimate the transit duration.  Many transit detection algorithms tend to discriminate against transit durations close to an alias of one day assuming the apparent transit event is likely caused by diurnal effects.  Planets with periods of $\sim$200 days would have durations close to 12 hours for F and G-type stars.  As will be seen in Sections \ref{indi} and \ref{sharingsection}, it is worth exploring those detections further, especially if only one or two events are observed during the year.  If a transit event has durations of several hours and is seen many times in a year it is likely caused by diurnal variations because short-period objects should have short transit durations.  The exceptions are a short-period planet orbiting an evolved star or one that is orbiting with high eccentricity and transiting near apastron, see Section \ref{assump} for details.  Care should be taken to keep those long-duration events that only appear a few times or else one risks throwing away a potential transiting, intermediate-period planet.  

The ease of searching for intermediate-period transiting planets is sometimes complicated by the requirement that photometry be calibrated over the entire baseline of the observations relative to a standard reference rather than only within a single night relative to neighbouring stars.  Rather than performing relative photometry using arbitrary field stars with similar colours and magnitudes of the target star on a night-to-night basis, the photometry will need to be referenced to a standard zero-point for all nights with observations, such as TYCHO or 2MASS stars located in the field.  Before addressing the question of detecting the transit signal in the data and making the effort to perform the necessary photometric calibrations, it must first be shown that transit groups have sufficient observing coverage to be observing a star while an intermediate-period transit is occurring in the first place.  We also investigate what the prospects of collaboration can do to improve those results.

\section{Numerical Technique}
\label{code}
The software is written in IDL and reads in a data file containing the Heliocentric Julian Dates (HJDs) that represent the times of observations of a particular field or generates artificial observing times.  For this paper we limit our analysis to one calendar year (2004) although it is easy to take into account multiple observing seasons.  It is assumed that the planet is in a circular orbit (see Section \ref{assump} for a discussion on the effects of eccentricity) and that the line-of-sight orbital inclination is 90$^\circ$ (a reasonable assumption since the planet is a transiting planet).  With these assumptions, the duration for a transit can then be expressed as \citep{mou2006}:
\begin{equation}
D_t \sim\frac{P R_*(1+\frac{R_p}{R_*})}{\pi a}
\label{durationest}
\end{equation}
where $P$ is the period, $R_*$ is the star's radius, $R_p$ is the planet's radius and $a$ is the orbital radius, and the approximation $\arcsin(x) \sim$$x$ is employed (compare to the full version in Eqn. \ref{guesspeq}).  The orbital radius is a function of the planet's period, the stellar mass and the planet's mass via Kepler's Third Law (Eqn. \ref{k3leq}).  Solving Eqn. \ref{k3leq} for $a$ and substituting into Eqn. \ref{durationest} yields:
\begin{equation}
D_t = \frac{P R_* (1+\frac{R_p}{R_*})}{\left(\frac{\pi G P^{2} (M_*+M_p)}{4}\right)^{1/3}}
\label{durationmaster}
\end{equation}
We assume for the purposes of this paper that the planet's radius is 1 Jupiter radius and the planet's mass is 1 Jupiter mass which we take to be 71492 km. and 1.8987x$10^{27}$ kg. respectively.  In that case Eqn. \ref{durationmaster} is a function of only three unknowns:  stellar mass, stellar radius and the planet's orbital period.  Canonical values are adopted for the stellar radii and masses from Table \ref{starprop}.  We only study spectral types F5 through M8 from Table \ref{starprop} for two main reasons.  First, the transit depth, and therefore the detectability by transit surveys, depends on $\left({\frac{R_p}{R_*}}\right)^{2}$, and $R_*$ grows much more rapidly for massive stars than $R_p$ for massive planets.  Second, stars above spectral type F5 have fewer spectroscopic lines in the optical wavelength regime for measuring precise radial velocities with optical spectrographs.

We assume a uniform extrasolar planet period distribution, which is inconsistent with observations.  Therefore, these numerical results are the fraction of transits observed when integrating over all possible phases \emph{if} an extrasolar planet existed with a given period.  It does not attempt to estimate the actual yield of wide-field transit surveys.  Section \ref{assump} briefly discusses binary star and extrasolar planet period distributions in the range explored by this paper.  The periods explored in this paper are taken in steps given by:
\begin{equation}
\Delta P \sim \frac{1}{24} exp\left({\frac{P-1}{167.}}\right)
\label{perstep}
\end{equation}
where $\Delta P$ is the step in period space, and $1\le$$P$$\le730$ days is the period.  Equation \ref{perstep} was chosen so that the minimum period step is $\sim$1 hour and the maximum period step is $\sim$3 days.  For each period, the phase is varied from 0 to 2$\pi$ in steps such that the difference in the start of each transit is 10 min.  Because the typical observing cadence for a wide-field transit survey is $\sim$10 min., this represents effectively all possible phases, or equivalently, all possible starting times for the transit.  An ``observation'' is defined as at least 10 epochs for which a photometry measurement was taken during a transit.  If multiple cameras observe during the same epoch, we still only consider that as one epoch.  This number was chosen since a typical minimum transit duration is $\sim$2 hours (for our simulation the minimum transit duration is $\sim$2.275 hours).  Requiring 10 data points with an assumed cadence of 10 minutes corresponds to observing $\sim$1.67 hours out of a $\sim$2 hour transit duration, which we assume is sufficient to qualify as an observation.  For reference, \citet{pon2006} found that a signal-to-noise of 8 is typically sufficient for a detection.

A better approach would be to model the data and calculate the signal-to-noise-ratios for each target, but to make a robust criteria would require white noise models and red noise models.  Typical values of red noise for long-period transit durations is unknown because current transit groups do not actively search for long-period transits and the magnitude of red noise increases as transit duration increases.  Making a simplistic assumption about white and red noise for such a diverse range of parent star types and transit durations is less robust than specifying a minimum ``covering fraction'' which we do.  This minimum of 10 data points is applied for all durations longer than 2 hours under the assumption that if a short-period transit of 2 hours can be detected than a transit with a longer duration and at least 2 hours of observations should also be detected.  It is worth emphasising that we are only calculating how many transits were \emph{observed} not necessarily detected, since most detection algorithms operate on folded data to improve the signal-to-noise while our interest is in how many of the transits were actually seen.  If, for example, a transiting planet in one month has 5 epochs observed during a transit and another 5 epochs observed during a transit in a second month, neither of those transits would count as being \emph{observed}, whereas a transit detection algorithm could fold the data to have a total of 10 epochs during the transit.

The number of transits seen for each period/phase combination is recorded and placed in one of four groups:  observed zero times, observed exactly one time, observed exactly two times, or observed three or more times.  Each of these groups is divided by the total number of phases sampled to determine the fraction of transits for each period that were observed zero, once, twice or multiple times.  In this way we approximately integrate over phase.  In this paper, we assume that any given transit event is actually a transiting planet, and not some other source of false-positive such as blended or grazing eclipsing binaries.  We also make no assumptions about the quality of each photometric data point or the ability for a group to detect the transit event.  That is, these probabilities are purely a function of times of observation.

\begin{figure}
\includegraphics[width=84mm]{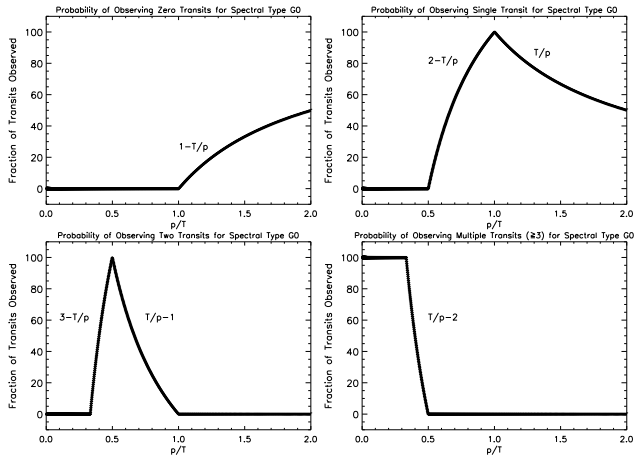}
\caption{Results from a trial run using 100$\%$ continuous observations (i.e. zero observing gaps) over 365 days, showing the fraction (in $\%$) of artificial planets observed zero times, exactly one time, exactly two times, and three or more times, as a function of period.  Looking at the plot for single transit events (top right), the expected peak at $p=T$ is recovered.  Planets with periods shorter than this value can be observed two or more times (for some phases) over the 365 day range, while planets with longer periods will have some phases where the transit occurs outside the 365 day range.  A similar expectation is met for the double transit observations (bottom left) and the peak at $p=\frac{T}{2}$.}
\label{space}
\end{figure}

\begin{table*}
\caption{\small{Analytical solutions for the continuous observation case.  $P$ is the orbital period and $T$ is the baseline of the observations (in this paper, $1 \le P \le 730$ days and $T=365$ days).  Columns are the probability of observing zero transits, exactly one transit, exactly two transits, or more than two transits.}\label{probspacetablecap}}
\centering
\begin{tabular}{@{}l l l l}
\hline
P(0) & P(1) & P(2) & P($\ge$3) \\
\hline
0 $\mid$ $P \leq T$ &  0 $\mid$ $P \leq \frac{T}{2}$   &  0 $\mid$ $P \leq \frac{T}{3}$  & 1 $\mid$ $P \leq \frac{T}{3}$ \\
\\
$1-\frac{T}{P}$ $\mid$ $P > T$  &  $2-\frac{T}{P}$ $\mid$ $\frac{T}{2} < P \leq T$  &  $3-\frac{T}{P}$ $\mid$ $\frac{T}{3} < P \leq \frac{T}{2}$  & $\frac{T}{P}-2$ $\mid$ $\frac{T}{3} < P \leq \frac{T}{2}$ \\
\\
        &  $\frac{T}{P}$ $\mid$ $P > T$  &  $\frac{T}{P}-1$ $\mid$ $\frac{T}{2} < P \leq T$   & 0 $\mid$ $P > \frac{T}{2}$ \\
\\
        &    &  0 $\mid$ $P > T$  & \\
\hline
\label{probspacetable}
\end{tabular}
\end{table*}

For verification and validation of the software, a case of continuous observations for 365 days (i.e. zero time lost) was used.  This could also be an extremely optimistic example of a space-based telescope where day-night cycles and weather effects do not hinder observations.  Fig. \ref{space} shows the results, where the fraction of artificial planets observed zero times, exactly one time, exactly two times and three or more times is shown as a function of period.  Table \ref{probspacetable} shows the analytical solutions for each of those cases, where $T$ is the overall baseline of observations (in this paper, $T=365$ days).  For reference, the non-trivial segments are labelled in Fig. \ref{space}.  As expected, the sum of all four plots yields $100\%$ for all periods.  The zero-observation plot (top left) has the expected shape where all planets with periods less than $p=T$ are observed at least once, while planets with periods greater than $p=T$ begin to have some phases where the transit happens outside the observational baseline explored.  For periods of $p=2T$ the fraction observed zero times is precisely 50$\%$.  For single transit events (top right) the expected peak of 100$\%$ is recovered at $p=T$.  Planets with periods less than this have the potential to be observed at least two times while planets with periods longer than this can occur outside the 365 day range explored.  It should be mentioned that Figs. \ref{space}, \ref{xoindiresults}, \ref{swindiresults}, \ref{eighthourplot}, \ref{sixteenhourplot}, \ref{esdiffsplot}, \ref{bothresults}, \ref{xodiffs} and \ref{swdiffs} assume the parent star is a G0 main-sequence star.  This is chosen as a representative case.  The (moderate) differences in assumption of spectral type are demonstrated in Fig. \ref{f5m5diffs} that compares an F5 to an M5 star.

\section{Results From Two Projects Individually}
\label{indi}
\begin{figure}
\includegraphics[width=84mm]{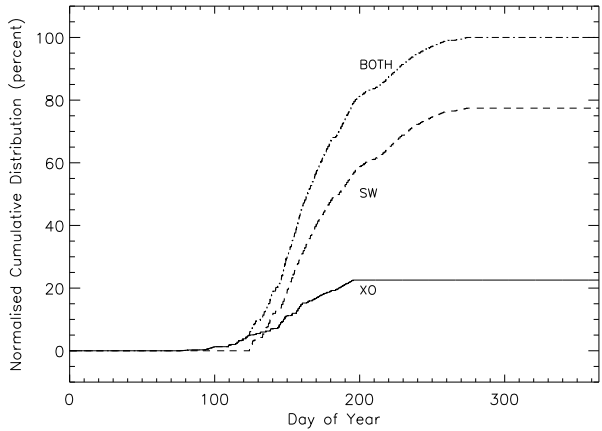}
\caption{Cumulative histogram of the times of observation for the XO Project and SuperWASP Project for a star located at RA = 16 02 42, DEC = +29 08 50 during the calendar year 2004.  The solid line is from the XO Project, dashed line is from the SuperWASP Project, and the dot-dashed line is the combined data sets.  Histogram is normalised to the total number of unique observations taken by XO and SuperWASP combined.  As can be seen, the XO Project contributes the most to the observations early on in the calendar year (Day 85-135), however SuperWASP quickly catches up and dominates the number of observations when it enters its period of intense observations starting near Day 130.}
\label{MASTERtimes}
\end{figure}

Given the dates of observations for a particular star the chances that a single transit event is actually a long-period companion or a short-period companion for which additional transits were missed can be determined.  To address this question and also to investigate the effects of collaboration amongst projects (Section \ref{sharing}), we desire a coordinate on the sky that a.) can be observed for a reasonably long timeframe by the projects involved (2-3 months at minimum) and b.) has a reasonable number of observations during that timeframe and approximates each project's observing strategy well (i.e., no 1-month gaps due to rare mechanical failures, etc.).  For this paper, we make use of data from the XO Project \citep{mcc2005} and the SuperWASP Project \citep{kan2007}.  We select a field on the basis of the above arguments with centre coordinate RA = 16 02, DEC = +29 09.  Fig. \ref{MASTERtimes} shows a cumulative histogram of the times of observations for this target by the XO Project and the SuperWASP Project during the 2004 calendar year.  The times of observations were binned into 5-minute intervals, yielding 1890 unique epochs for XO and 4572 unique epochs for SuperWASP.  If multiple cameras from the same group observed during the same 5-minute bin, only one was counted, i.e., the data were separated into boolean bins.  The solid line is from the XO Project, the dashed line is from the SuperWASP Project, and the dot-dashed line is the combined data sets.  The histogram is normalised to the total number of unique observations taken by XO and SuperWASP combined.

As can be seen, the XO Project contributes the most to the observations early on in the calendar year (Day 85-135), however SuperWASP quickly dominates the number of observations when it enters its period of intense observations starting near Day 130.  SuperWASP-N is located at La Palma in the Canary Islands while XO is located on Mt. Haleakala in Hawaii so they make an excellent pair for exploring collaboration between multi-longitudinal observatories.  The observing strategies of XO and SuperWASP are different.  While XO limits a given field to a maximum number of hours per night, SuperWASP generally observes a given field as often as possible per night.  This particular coordinate during the 2004 calendar year was observed by SuperWASP with multiple cameras per night because it was near a boundary of a field, which explains the greater number of epochs.  In later calendar years, when SuperWASP had all of its cameras operating, this field was only observed by one camera and the number of epochs per night for XO and SuperWASP are quite similar.  We use this particular field and observing season to demonstrate the advantages of collaboration even if a given project has more observations of a given field than another.  Artificial cases presented in Section \ref{eightsixteen} provide an upper limit on the advantages of collaborating when the two projects observe exactly the same number of epochs.  While the SuperWASP observations for this particular object have a greater density, the XO Project provides increased coverage during the overlap region as well as observations taken earlier in the year before the SuperWASP project started to observe this target.

It should be noted that a ``typical'' observing window is difficult to define in this study.  The results depend upon the duty cycle of the observations and the number, duration and location of observing gaps.  The theoretical, simulated observing windows presented in Sections \ref{code} and \ref{eightsixteen} are one type of upper limit, while the real observing windows we have chosen from XO and SuperWASP are used primarily as a demonstration of our numerical technique and one example of an application to real data.  The lack of extremely long gaps in observations due to critical errors for both projects makes this field an appropriate choice for a demonstration.

\begin{figure}
\includegraphics[width=84mm]{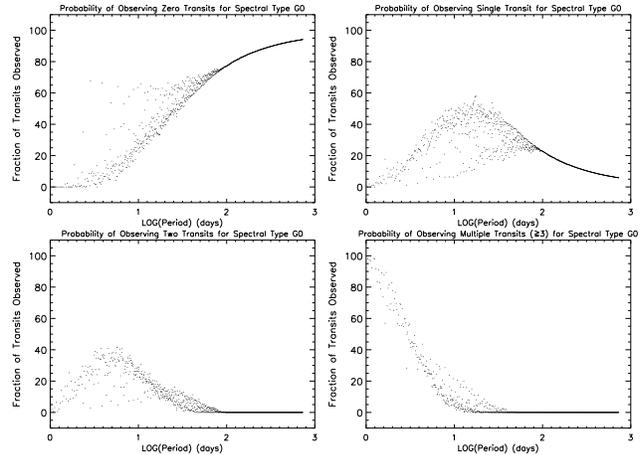}
\caption{Probability of observing zero, single, double and multiple transit events as a function of period for the XO Project based solely on times of observation for a 1.0 M$_{J}$ planet orbiting a typical G0 main sequence star.  Even with the XO Project's observations alone the chances of a single transit event being a companion with an intermediate-length period (between 14-100 days) is greater than it being a single transit event of a short-period Hot Jupiter (top right, $\sim$20-50$\%$ vs. $\sim<10\%$, assuming $i=90^\circ$, $e=0$ and a uniform semi-major axis distribution).  The probability of a double transit event being a companion with a period of a few weeks is also fairly high (bottom left).  As expected, chances of the XO Project observing multiple transits of Hot Jupiters is very high (bottom right).}
\label{xoindiresults}
\end{figure}

\begin{figure}
\includegraphics[width=84mm]{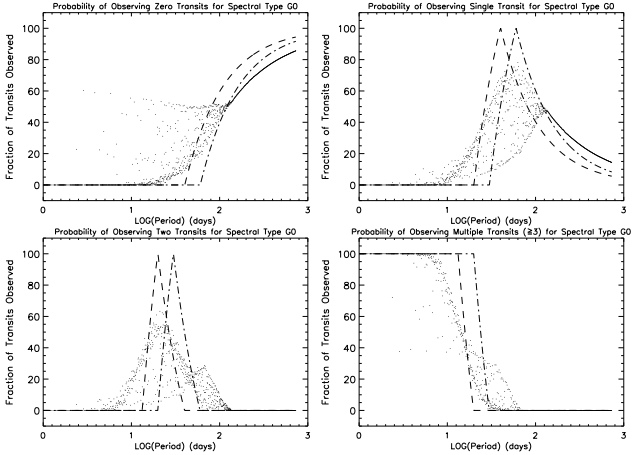}
\caption{Probability of observing zero, single, double and multiple transit events as a function of period for the SuperWASP Project based solely on times of observation for a 1.0 M$_{J}$ orbiting a typical G0 main sequence star.  The dashed line is for an artificial case of 40 days of continuous observations, while the dot-dashed is for 60 days of continuous observations, for comparison.  There is a significant probability that single transit events are of intermediate-period, particularly around 60 days, while it is highly unlikely they would be short-period Hot Jupiters only observed once (top right).  As expected, short-period planets have a very high probability of being observed three or more times (bottom right).}
\label{swindiresults}
\end{figure}

We use the probability software to estimate the chances that a single transit event might be a ``Temperate Jupiter'' for XO and SuperWASP, individually.  Fig. \ref{xoindiresults} shows the results for the XO Project, while Fig. \ref{swindiresults} shows the results for SuperWASP for companions orbiting a G0 main sequence star (see Fig. \ref{f5m5diffs} for the differences between assumed spectral types).  Results from Figs. \ref{xoindiresults} and \ref{swindiresults} using a bin size of $P=15$ days are presented in Table \ref{tabulatedresults1} in the Appendix.  First we examine the XO results (Fig. \ref{xoindiresults}).  As expected, the XO Project does not miss many Hot Jupiters (zero-transit observations, top-left).  However, for objects with periods greater than 30 days the probability of missing the transits is greater than 50$\%$.  Single transit events peak in probability near a period of 17 days (top right) while only the shortest-periods are observed two or more times (bottom left and bottom right).  The SuperWASP results (Fig. \ref{swindiresults}) are quite different.  In this figure, an artificial case of 40 days of continuous observations (dashed line) and 60 days of continuous observations (dot-dashed line) are overplotted for comparison.  The most striking effect is the existence of a double-peak in the single transit, double transit and multiple transit plots.  Upon closer examination similar peaks are present in the XO results, though much less well-defined.  Because of the density of the SuperWASP observations compared to XO, the probability of not observing any transits is lower for all periods explored.  Indeed, even for objects with periods out to 200 days the probability of observing at least one transit is $\sim$50$\%$ (top left) while single transit events peak at $\sim$60 days (almost 80$\%$).  Most likely this peak behaviour is caused by the observing strategy of SuperWASP, which concentrated observations of this target for approximately two months ($\sim$days 140-200 in Fig. \ref{MASTERtimes}).  We can compare this period of frequent observations with the space-based case (Fig. \ref{space}), where in the ground-based case the time lost is primarily due to daylight.  Planets with periods close to 60 days have a large probability of being seen once during this two month stretch of intense observations.  Planets with larger periods than this 2-month stretch of intense observations are essentially independent of its effects and indeed there is a smooth, continuous decrease in probability that follows the analytic solution.  Planets with periods less than this critical value of 60 days exhibit a wide range of effects depending on the precise value of the period due to daylight considerations.  Planets with extremely short periods (like Hot Jupiters) transit so often over the course of a year that the effects of this two month ``intense observation period'' are minimal in terms of single or double transit events because they are likely to be observed three or more times.  In the double-observation plot (bottom left) the peak at $\sim$60 days is again observed.  The stronger peak at shorter period occurs near 20 day periods, which most closely resembles a 40-day continuous, space-like case.  In addition, examination of the multiple transit plot (bottom-right) demonstrates how different the SuperWASP data over an entire year is compared to a short-term, continuous monitoring situation, particularly at periods between 20-40 days.  The two oscillating structures between $P$=100-130 days is because the period step at those values based on Eqn. \ref{perstep} is $\sim$0.5-0.6 days, and therefore it is due to how we step through period space and diurnal effects.

\section{The Advantage of Combining Data}
\label{sharingsection}
\subsection{Simulated Results}
\label{eightsixteen}
\begin{figure}
\includegraphics[width=84mm]{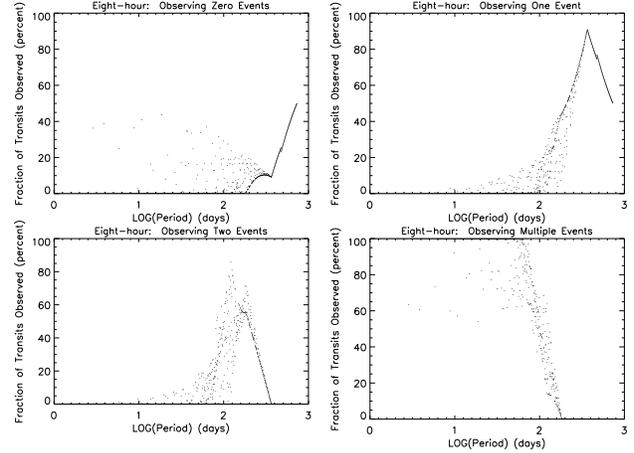}
\caption{Results using a simulated observing window consisting of eight hours of continuous observations every night for 365 days.  The scatter at short periods is due to the artificial day-night cycle.  A discontinuity occurs between a period of 473.5-475.5 days (top-right), where a transit would last $\sim$16.75 hours.  We expect this is also due to the day-night cycle, as it does not appear in the sixteen-hour plot (Fig. \ref{sixteenhourplot}) or the space-based case (Fig. \ref{space}).}
\label{eighthourplot}
\end{figure}

\begin{figure}
\includegraphics[width=84mm]{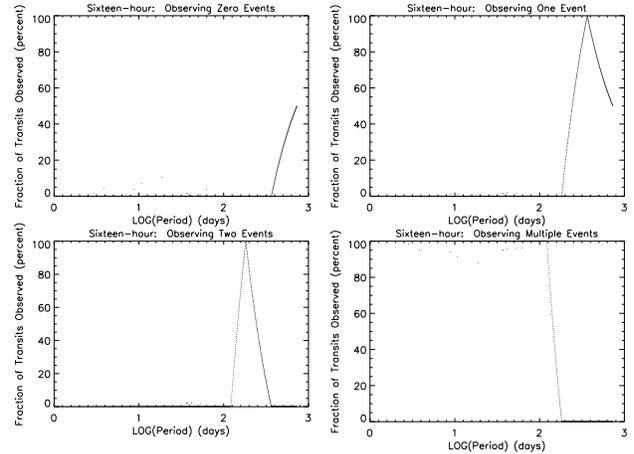}
\caption{Results using a simulated observing window consisting of sixteen hours of continuous observations every night for 365 days.  As can be seen, the scatter at short periods is nearly eliminated when a larger duty cycle is used.}
\label{sixteenhourplot}
\end{figure}

\begin{figure}
\includegraphics[width=84mm]{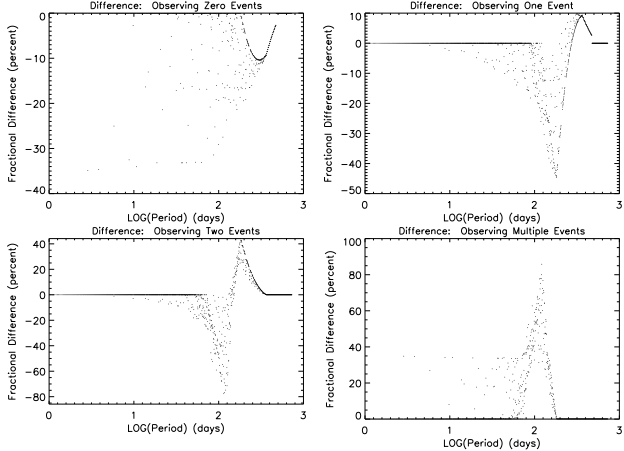}
\caption{Difference between the sixteen-hour case and the eight-hour case.  Results are most significant in the multiple-observation plot (bottom right) for periods between 50 and 150 days.  These results are approximate upper limits to the advantage of multi-longitudinal wide-field transit operations.  The discontinuity at $P\sim475$ days (top-right) is discussed in the caption to Fig. \ref{eighthourplot}.}
\label{esdiffsplot}
\end{figure}

Collaboration of wide-field transit projects allows for an increased chance at observing long-period transiting planets and reduces the probability of missing transits of short-period Hot Jupiters.  For rare events such as intermediate-period transiting planets, multiple telescopes independently observing the same single-transit event greatly increases the confidence that the event is real.  In addition, multi-longitudinal telescopes can increase the speed of follow-up observations via their year-round observing capability of targets, increasing the rate of discovery.  To investigate the maximum advantage of collaborating, we simulate a multi-longitudinal observing program.  In one case, we simulate observing continuously for eight hours every night for 365 days (Fig. \ref{eighthourplot}), which we define as an unrealistic, best-case scenario for a single site.  No seasonal effects such as the varying length of night were implemented.  To the extent that these simulations are similar to the 100$\%$ continuous observations (Fig. \ref{space}), the results are insensitive to the choice of $T=365$ days because these curves depend on $\frac{T}{P}$.  We then simulate observing continuously for sixteen hours every night for 365 days (Fig. \ref{sixteenhourplot}).  Data from Fig. \ref{eighthourplot} and Fig. \ref{sixteenhourplot} using a bin size of $P=15$ days are presented in Table \ref{tabulatedresults2} in the Appendix.  This simulates an ideal collaboration using the simplest observing windows and zero overlap between the two sites.  The scatter at short periods in Fig. \ref{eighthourplot} is caused by the artificial day-night cycle.  Due to the dense sampling of period-space, certain values for the period tend to have transits occurring during daylight.  This effect is greatly reduced for the sixteen-hour case.  A discontinuity occurs between a period of 473.5-475.5 days in the eight-hour case (top-right plot), where a transit would last $\sim$16.75 hours.  We expect this is also due to the day-night cycle, because it does not appear in the sixteen-hour plot (Fig. \ref{sixteenhourplot}) or the space-based case (Fig. \ref{space}).  The difference between the sixteen-hour case and the eight-hour case is plotted in Fig. \ref{esdiffsplot}.  The effect is most noticeable in the range 50$<p<$150 days, where up to 80$\%$ of companions are observed three or more times compared to the single-site case.  The effect of longer periods tends to be more modest, generally a 10-20$\%$ increase in single or double-transit observations.

\subsection{Results Using XO and SuperWASP Data}
\label{sharing}
\begin{figure}
\includegraphics[width=84mm]{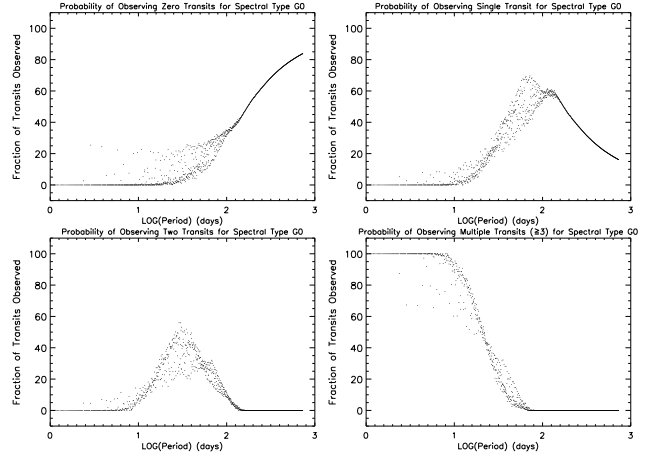}
\caption{Probability of observing zero, single, double and multiple transit events as a function of period for the combined SuperWASP+XO dates of observation.  Though qualitatively similar to the SuperWASP result, the addition of the XO data set does result in a decrease of single and double transit event Hot Jupiters (meaning better efficiency and a larger yield in Hot Jupiter discovery) as well as increased probabilities of observing long-period planets (greater chance at observing a transiting ``Temperate Jupiter'').}
\label{bothresults}
\end{figure}

\begin{figure}
\includegraphics[width=84mm]{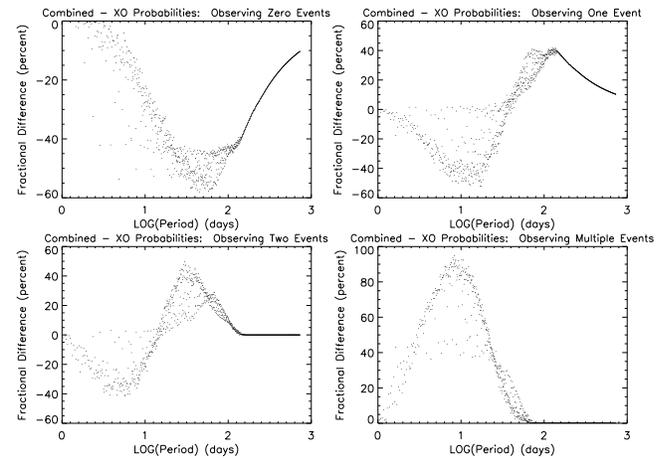}
\caption{Difference in probabilities when using the combined XO+SuperWASP data set vs. XO alone.  As can be seen, chances of the dreaded ``zero observations'' are reduced for almost all periods, and significantly reduced for intermediate-periods (e.g., 50 days, top left).  single transit events of long-period planets is increased, while single and double transit events of short-period Hot Jupiters is decreased (they become multiple-observations, which for the 14-50 day range are drastically improved, bottom right).}
\label{xodiffs}
\end{figure}

\begin{figure}
\includegraphics[width=84mm]{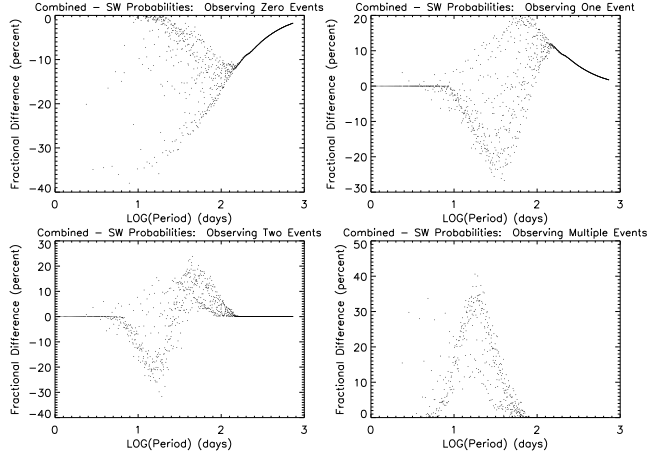}
\caption{Same as Fig. \ref{xodiffs} but for the SuperWASP only data vs. the combined data set.  The improvements are not quite as drastic yet still are of the 10-20 $\%$ level for intermediate-length periods.  For periods in the $\sim$25 day range the affect of adding XO's contribution is quite significant in obtaining multiple transit observations.}
\label{swdiffs}
\end{figure}

We perform the same probability calculations as done in Section \ref{indi} on the combined dates of observation for XO and SuperWASP.  This simulates a collaboration between the two groups where photometry is shared and calibrated to a standard reference such that the zero points were equivalent.  Alternatively, this represents the same group constructing two observatories at different longitudes to improve their observing coverage.  Fig. \ref{bothresults} shows the result, while the results using a bin size of $P=15$ days are presented in Table \ref{tabulatedresults1} in the Appendix.  Though qualitatively similar to the SuperWASP results, the addition of the XO Project increases the number of intermediate-period planets observed exactly once by reducing the number not observed at all (compare the top two graphs of Figs. \ref{swindiresults} and \ref{bothresults} at periods between 0-200 days).  Figs. \ref{xodiffs} and \ref{swdiffs} show the difference between using the combined XO+SuperWASP data and each of them individually.  The effect on the XO Project's results is significant.

As can be seen in Fig. \ref{xodiffs}, periods between $\sim$30 and 150 days have a $\sim$30-60$\%$ reduction in the number observed zero times (top left).  Planets with periods between $\sim$30-100 days have the number of single and double transit events increased by 20-40$\%$ (top right, bottom left).  The multiple-observation planets have their period range greatly increased, meaning if XO combined their data with SuperWASP they could extend the planets for which three or more transits are likely to be observed (i.e. probability $\sim$50$\%$) out to periods approaching 30 days.  From SuperWASP's perspective the effects are not as large but they are still significant (Fig. \ref{swdiffs}).  In particular, multiple-observation planets with periods out to 30 days increase by 20-40$\%$, increasing SuperWASP's chances of observing a transit with these periods.  Double transit observations for periods between 30-100 days increase by as much as $20\%$.  In this case, the advantages of collaboration between wide-field transit groups (or building a multi-longitudinal observing program) has its clear advantage.  The two oscillating patterns for periods $P$=100-130 days is once again caused by how we step through period space (at those periods, $dP$=0.5-0.6 days, so it is a diurnal effect).

\begin{figure}
\includegraphics[width=84mm]{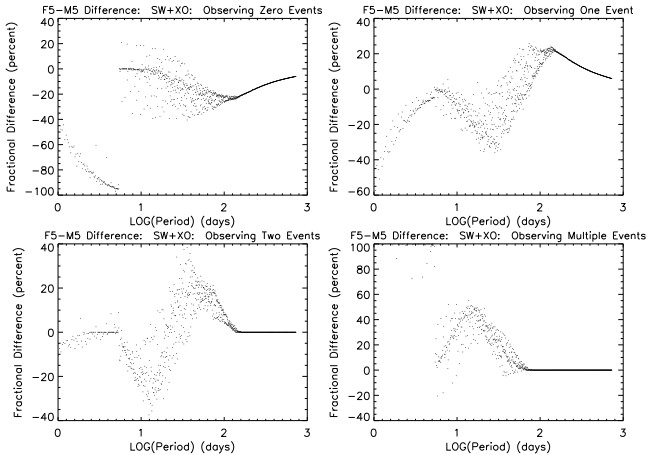}
\caption{Difference between observing zero, exactly one, exactly two or multiple transit events for F5 compared to M5 stars, assuming a circular orbit and a 1.0 M$_{J}$ planet, using the combined XO and SuperWASP times of observation.  For M-type stars where the Habitable Zone is on the order of 1-2 months, it is up to $\sim$30$\%$ less likely that such a planet would be observed three or more times than for an F-type star and similar periods.  Single or double transit events should be given even more consideration, since planets with 1-2 month periods are more likely to be observed only once or twice around M-type stars compared to more massive stars.}
\label{f5m5diffs}
\end{figure}

At this point we investigate how the spectral type, and hence stellar radius and mass, affects the results.  The most extreme comparison would be to compare the F5 and M5 results (M8 stars are fairly rare in the apparent magnitude range that most wide-field transit surveys operate).  Fig. \ref{f5m5diffs} shows the difference in observation probability between an F5 and M5 star using the combined XO and SuperWASP times of observation.  Results using a bin size of $P=15$ days are presented in Table \ref{tabulatedresults2} in the Appendix.  The main differences are the size of the stellar disks and the orbital radius of the planet (for a fixed planet mass, a given period is located farther away for a more massive star than a less massive star).  Both these effects serve to change the duration of a transit.  Chances of observing zero transits are greater for M stars than more massive stars for the entire period range explored (generally 15-20$\%$).  Multiple observations (three or more) of planets with periods of $\sim$15 days is 50$\%$ greater around more massive stars than M-type stars.  However, it is more likely that such planets would be observed exactly twice (bottom left) around M-type stars ($\sim$40$\%$).  Since the Habitable Zone of M-type stars is on the order of 1-2 months (for spectral types M0-M5) this is a period regime of particular interest.  Our results suggest that for M-type stars, it is likely that Habitable Zone planets will be observed only once around M-type stars (top right, up to a 30$\%$ increased chance to detect exactly once around an M5 compared to an F5), and therefore such observations of a single transit event merits even further consideration if the star is determined to be an M-type star, as would be evident from its proper motion and colour.  Dedicated searches like the MEarth Project \citep{nut2007} that have nearly continuous observations should detect more than a single transit.  For current wide-field transit searches, the lack of M-type stars in the magnitude range being searched is another important factor in the expected yield of transiting planets around M-type stars.  It is estimated that the magnitude limit must extend beyond $V$=16 before a statistically significant population of transiting planets around M-type stars can be expected \citep{mcc2007}.

\section{Effects of Eccentricity and Binary Star Contamination}
\label{assump}
The results provided above are based on several assumptions.  We assume a circular orbit for all planets.  We also make no claims about how many objects are actually extrasolar planets as opposed to astronomical false positives such as grazing or diluted eclipsing binaries.  We also assume that all photometric data points carry equal weighting.  We make no claims about the ability to pick out these single and double transit events from actual data because it is dependent on the period-finding algorithm employed by the group and the signal-to-noise of the data.  In this section we will discuss some of the effects of these assumptions on the conclusions presented above.

Regarding the assumption of eccentricity it is well known that extrasolar planets have a much wider range of eccentricities than our own system.  The two primary effects of eccentricity are on the geometric transit probability and the duration of a transit.  \citet{bar2007} shows that the geometric probability of a transit actually increases by a factor of $(1-e^2)^{-1}$ for eccentric planets.  He points out that approximately 12$\%$ of known extrasolar planets with radial velocity measurements have eccentricities greater than 0.5 and nearly 50$\%$ have eccentricities greater than Mercury's (0.2056), which is the most eccentric planet in the Solar System.  The intermediate-period planets discussed in this paper all have orbital distances much larger than the tidal circularisation regime, so these intermediate-period planets can have a range of eccentricities.  The geometric probabilities presented in Table \ref{probtable} are therefore \emph{lower limits} because planets are more likely to transit near periastron and with greater geometric transit probabilities compared to the circular case.

\citet{bar2007} also showed that an eccentric planet travels with a speed that is a factor of $\sqrt{\frac{1+e}{1-e}}$ faster at periapsis and a factor of $\sqrt{\frac{1-e}{1+e}}$ slower at apoapsis compared to a planet in a circular orbit with the same semi-major axis.  Following \citet{mou2006} the duration of a transit including eccentricity can be expressed as:
\begin{equation}
D_t = 2 \sqrt{1-\left(\frac{\rho \cos(i)}{R_*+R_p}\right)^2} (R_*+R_p) \frac{\sqrt{1-e^2}}{1+e \cos(\phi)} \left(\frac{P}{2 \pi GM_*}\right)^{1/3}
\label{durationecc}
\end{equation}
where $\rho$ is the star-planet distance at the time of transit, $\phi$ is the orbital phase at the time of transit and $i$ is once again the line-of-sight orbital inclination.  For a planet in a circular orbit with the same period compared to an eccentric planet at $i=90^\circ$, the difference in transit duration is $\frac{\sqrt{1-e^2}}{1+e \cos(\phi)}$.  For periapsis we define $\phi = 0^\circ$ and for apoapsis $\phi = 180^\circ$.  For a planet at periapsis, an eccentricity of approximately 0.6 will result in a transit duration that is 50$\%$ shorter.  In the case of the intermediate-period planets that are the primary focus of this paper, this reduction is beneficial.

As seen in Fig. \ref{durations} the transit duration for planets with periods close to 200 days is nearly 12 hours.  Because of the methods used by most wide-field transit groups, targets that exhibit photometric variability with periods close to an alias of 24 hours are ignored and assumed to be diurnal variation.  An eccentric planet near periapsis will shorten the transit duration from an alias of 24 hours, and, for extremely eccentric planets, could shorten the duration to be only a few hours, precisely the duration of what the transit groups are looking for (i.e. hot Jupiters).  Eccentric planets near apoapsis have the opposite effect and increase their transit duration relative to a circular orbit with the same semi-major axis.  Aside from the difficulties of detecting planets with very long durations, an increase in transit duration will only increase the probability of observing the transit.  In terms of the definition of an ``observation'' in this paper, an increase in transit duration can only increase the probability of observing the planet.

A planet near apoapsis is less likely to transit its parent star compared to a planet in periapsis (for an eccentricity of 0.5, the geometric transit probability of an apoapsis planet is a factor of 3 lower than a periapsis planet).  Since the argument of periastron is a uniform distribution on the sky, and eccentric planets closest to periastron are most likely to transit their parent stars, the distribution of transiting eccentric planets should be skewed towards those transiting near periapsis.  Therefore the general effect of eccentricity on transiting extrasolar planets should be to \emph{reduce} the duration of a transit.  In terms of actually detecting the transits from the data, the general effect of eccentricity should be to create shorter-duration transits that are closer to the Hot Jupiter transit durations currently being sought.

Grazing eclipsing binary stars and blended binaries are two of the most common astronomical sources that mask themselves as transiting hot Jupiters \citep{bro2003}.  Although effort is made to reduce the number of false detections due to these sources or determine the binary nature of targets before conducting follow-up, it is fairly often the case that the binary star nature of transiting planet candidates can only be determined via radial velocity measurements.  The binary period distribution is consistent with a distribution that is uniform in log space for $1<\log(P)<3$ (where $P$ is measured in days) \citep[][based on work by \citet{duq1991,hal2003}]{egg2004}, while the known extrasolar gas/ice giant planet distribution is more uncertain, although the distribution can be described as a ``pile-up'' near periods of $\sim 3$ days and an increasing distribution at longer periods \citep{udr2007}.

Interesting structure in the intermediate-period regime such as a bimodal ``pileup'' between periods of ~3 days and ~300 days has been suggested \citep{udr2003,bro2003} and may be a boundary region between two different migration regimes.  \citet{bro2003} estimates that for objects with periods between 1 - 30 days, typical planet transit depths (between 1 and 5 millimags) and a sample of 10,000 stars that 0.0953$\%$ of the stars will have at least one transit with those types of depths.  Out of those, 48$\%$ are grazing eclipsing binaries and 37$\%$ are some kind of diluted eclipsing binary, the rest being planetary companions (15$\%$), although it should be noted that he does not carry out his analysis to longer periods.  Qualitatively though, we can say that in the period range of $\sim$10 to $\sim$300 days where there is a relative lack of extrasolar gas/ice giant planets, binary star contamination might be greater.

\section{Summary}
\label{summary}
Transiting intermediate-period planets offer many exciting research opportunities, but require photometry that is calibrated over the course of an entire observing program to a standard reference in the field, rather than calibrated on a night-to-night basis relative to neighbouring stars.  Many wide-field transit surveys have targets for which only single or double transit events were observed.  These transits could be due to instrumental effects, short-period transits for which additional transits were missed or intermediate-period transiting planets.  A necessary condition before attempting to modify the data reduction procedures on the massive amounts of data that wide-field transit surveys have now collected in an attempt to search for these objects is to verify that their observing strategies are sufficient to \emph{observe} these targets solely as a function of observing times.  Towards that end we have described a method that will calculate the number of transits observed for various periods as a function of observing times by effectively integrating over all possible phases.  We have used data from the 2004 observations of the XO Project and SuperWASP Projects to investigate what the observing probabilities are independently and combined.

While the SuperWASP Project has a greater density of observations for the field selected, we see that both the XO Project and SuperWASP can benefit greatly by collaborating.  For the period range $14 < P < 150$ days, both groups experience an increase in single and double transit events up to 20-40$\%$ (Figs. \ref{xodiffs}, \ref{swdiffs}).  For the XO Project using its data alone,  $\sim$20-40$\%$ of planets with periods 14-150 days should have been observed at least once (Fig. \ref{xoindiresults} top-left).  For the SuperWASP Project, $\sim$50-90$\%$ of planets with periods between 14-150 days should have been observed at least once (Fig. \ref{swindiresults} top-left).  If XO and SuperWASP combined their observations, 50-100$\%$ of planets with periods less than 20 days should be observed three or more times.  Results depend upon the observing window of the field, including the number, length and positions of gaps in the observations.

We find that in general wide-field transit surveys have selected appropriate observing strategies to observe a significant fraction of transiting giant planets with semimajor axes larger than the Hot Jupiter regime.  The actual number of intermediate-period transiting planets that are detected depends upon their true semimajor axis distribution and the signal-to-noise of the data.  We therefore conclude that the investment of resources needed to investigate more sophisticated photometry calibrations or the resources needed to examine single and double transit events from wide-field surveys might be a worthwhile endeavour.Whether these events are from instrumental sources, astronomical false positives, or actual planets, understanding the origins of these single and double transit events is an important task to improve data quality and reliability, improve false-alarm rejection criteria, or discover planets which can offer some of the first direct comparisons of the radii, densities, and atmospheric compositions of extrasolar gas giants in environments that more closely resemble our own.

\section{Acknowledgements}
\label{ack}
We thank the referee, Frederic Pont, for a quick response and several suggestions that improved the quality of the paper.  The authors thank Justin Crepp, Suvrath Mahadevan, Dimitri Veras and Eric Ford for useful discussions.  S.W.F. acknowledges support from the Florida Space Grant Fellowship and the Space Telescope Science Institute Summer Student Program.  P.R.M. acknowledges NASA Origins grant NAG5-13130.  This publication has made use of the SIMBAD database, operated at CDS, Strasbourg, France, the Extrasolar Planets Encyclopedia maintained by Jean Schneider (http://exoplanet.eu/index.php) and the California-Carnegie Catalog of Nearby Exoplanets (http://exoplanets.org/planets.shtml).

\appendix
\section{Tabulated Results}
We present tabulated results from Figs. \ref{xoindiresults}, \ref{swindiresults} and \ref{bothresults} (Table \ref{tabulatedresults1}), and Figs. \ref{eighthourplot}, \ref{sixteenhourplot} and \ref{f5m5diffs} (Table \ref{tabulatedresults2}).  We bin the results using a bin size of $P$=15 days and take the mean of all results within each bin.  The primary motivation is to mitigate the effects of daylight in the short period regime, which causes the ``chaotic'' behaviour seen in, e.g., Fig. \ref{bothresults}.  This behaviour is caused by the day-night cycle acting on the small differences (of order hours) in the periods being explored.
\begin{table*}
\centering
\caption{~Tabulated results from XO (Fig. \ref{xoindiresults}), SuperWASP (Fig. ref{swindiresults}) and XO+SuperWASP (Fig. \ref{bothresults}), using a bin size of 15 days and taking the median of all results within that bin.  Columns 1 and 2 are the bin range.  P(0), P(1), P(2) and P($\ge$3) are the probabilities of observing zero transits, a single-transit, a double-transit or a multi-transit for planets with those periods, given in percent ($\%$).  Table is available in electronic format upon request.}
\begin{tabular}{@{}l l l l l l l l l l l l l l}
\hline
\small{$P_{min}$} & \small{$P_{max}$} & \small{XO} & & & & \small{SW} & & & & \small{XO+SW} & & &
\\
   \small{(days)}     &    \small{(days)}     & \small{P(0)} & \small{P(1)} & \small{P(2)} & \small{P($\ge$3)} & \small{P(0)} & \small{P(1)} & \small{P(2)} & \small{P($\ge$3)} & \small{P(0)} & \small{P(1)} & \small{P(2)} & \small{P($\ge$3)}
\\
\hline

 \small  0 &  \small 15 &  \small 19.81 &  \small 27.66 &  \small 23.78 &  \small 28.75 &  \small  4.09 &  \small  2.94 &  \small  9.19 &  \small 83.79 &  \small  1.33 &  \small  1.50 &  \small  4.43 &  \small 92.75
          \\
 \small 15 &  \small 30 &  \small 48.26 &  \small 38.90 &  \small 11.15 &  \small  1.69 &  \small 11.98 &  \small 24.71 &  \small 39.23 &  \small 24.08 &  \small  4.68 &  \small 14.88 &  \small 31.99 &  \small 48.45
          \\
 \small 30 &  \small 45 &  \small 60.75 &  \small 33.49 &  \small  5.47 &  \small  0.28 &  \small 21.14 &  \small 43.77 &  \small 27.65 &  \small  7.45 &  \small 10.26 &  \small 34.85 &  \small 37.64 &  \small 17.25
          \\
 \small 45 &  \small 60 &  \small 66.99 &  \small 30.38 &  \small  2.61 &  \small  0.01 &  \small 29.47 &  \small 48.75 &  \small 18.88 &  \small  2.90 &  \small 16.01 &  \small 47.83 &  \small 30.05 &  \small  6.12
          \\
 \small 60 &  \small 75 &  \small 71.62 &  \small 26.87 &  \small  1.50 &  \small  0.00 &  \small 36.88 &  \small 47.69 &  \small 15.08 &  \small  0.35 &  \small 22.38 &  \small 52.64 &  \small 23.75 &  \small  1.23
          \\
 \small 75 &  \small 90 &  \small 74.43 &  \small 25.14 &  \small  0.44 &  \small  0.00 &  \small 42.61 &  \small 46.70 &  \small 10.69 &  \small  0.00 &  \small 27.66 &  \small 55.76 &  \small 16.55 &  \small  0.02
          \\
 \small 90 &  \small105 &  \small 76.83 &  \small 23.12 &  \small  0.05 &  \small  0.00 &  \small 46.67 &  \small 46.20 &  \small  7.14 &  \small  0.00 &  \small 32.44 &  \small 56.76 &  \small 10.79 &  \small  0.00
          \\
 \small105 &  \small120 &  \small 79.00 &  \small 21.00 &  \small  0.00 &  \small  0.00 &  \small 48.89 &  \small 47.83 &  \small  3.29 &  \small  0.00 &  \small 35.75 &  \small 58.44 &  \small  5.81 &  \small  0.00
          \\
 \small120 &  \small135 &  \small 80.77 &  \small 19.23 &  \small  0.00 &  \small  0.00 &  \small 51.42 &  \small 47.55 &  \small  1.03 &  \small  0.00 &  \small 38.98 &  \small 58.53 &  \small  2.48 &  \small  0.00
          \\
 \small135 &  \small150 &  \small 82.22 &  \small 17.78 &  \small  0.00 &  \small  0.00 &  \small 54.19 &  \small 45.81 &  \small  0.00 &  \small  0.00 &  \small 42.33 &  \small 57.05 &  \small  0.62 &  \small  0.00
          \\
 \small150 &  \small165 &  \small 83.39 &  \small 16.61 &  \small  0.00 &  \small  0.00 &  \small 57.50 &  \small 42.50 &  \small  0.00 &  \small  0.00 &  \small 46.23 &  \small 53.70 &  \small  0.07 &  \small  0.00
          \\
 \small165 &  \small180 &  \small 84.46 &  \small 15.54 &  \small  0.00 &  \small  0.00 &  \small 60.22 &  \small 39.78 &  \small  0.00 &  \small  0.00 &  \small 49.86 &  \small 50.14 &  \small  0.00 &  \small  0.00
          \\
 \small180 &  \small195 &  \small 85.36 &  \small 14.64 &  \small  0.00 &  \small  0.00 &  \small 62.53 &  \small 37.47 &  \small  0.00 &  \small  0.00 &  \small 53.03 &  \small 46.97 &  \small  0.00 &  \small  0.00
          \\
 \small195 &  \small210 &  \small 86.02 &  \small 13.98 &  \small  0.00 &  \small  0.00 &  \small 64.55 &  \small 35.45 &  \small  0.00 &  \small  0.00 &  \small 55.60 &  \small 44.40 &  \small  0.00 &  \small  0.00
          \\
 \small210 &  \small225 &  \small 86.69 &  \small 13.31 &  \small  0.00 &  \small  0.00 &  \small 66.42 &  \small 33.58 &  \small  0.00 &  \small  0.00 &  \small 57.87 &  \small 42.13 &  \small  0.00 &  \small  0.00
          \\
 \small225 &  \small240 &  \small 87.30 &  \small 12.70 &  \small  0.00 &  \small  0.00 &  \small 67.88 &  \small 32.12 &  \small  0.00 &  \small  0.00 &  \small 59.79 &  \small 40.21 &  \small  0.00 &  \small  0.00
          \\
 \small240 &  \small255 &  \small 87.81 &  \small 12.19 &  \small  0.00 &  \small  0.00 &  \small 69.29 &  \small 30.71 &  \small  0.00 &  \small  0.00 &  \small 61.69 &  \small 38.31 &  \small  0.00 &  \small  0.00
          \\
 \small255 &  \small270 &  \small 88.30 &  \small 11.70 &  \small  0.00 &  \small  0.00 &  \small 70.49 &  \small 29.51 &  \small  0.00 &  \small  0.00 &  \small 63.39 &  \small 36.61 &  \small  0.00 &  \small  0.00
          \\
 \small270 &  \small285 &  \small 88.74 &  \small 11.26 &  \small  0.00 &  \small  0.00 &  \small 71.62 &  \small 28.38 &  \small  0.00 &  \small  0.00 &  \small 64.97 &  \small 35.03 &  \small  0.00 &  \small  0.00
          \\
 \small285 &  \small300 &  \small 89.13 &  \small 10.87 &  \small  0.00 &  \small  0.00 &  \small 72.65 &  \small 27.35 &  \small  0.00 &  \small  0.00 &  \small 66.39 &  \small 33.61 &  \small  0.00 &  \small  0.00
          \\
 \small300 &  \small315 &  \small 89.50 &  \small 10.50 &  \small  0.00 &  \small  0.00 &  \small 73.60 &  \small 26.40 &  \small  0.00 &  \small  0.00 &  \small 67.71 &  \small 32.29 &  \small  0.00 &  \small  0.00
          \\
 \small315 &  \small330 &  \small 89.82 &  \small 10.18 &  \small  0.00 &  \small  0.00 &  \small 74.47 &  \small 25.53 &  \small  0.00 &  \small  0.00 &  \small 68.89 &  \small 31.11 &  \small  0.00 &  \small  0.00
          \\
 \small330 &  \small345 &  \small 90.13 &  \small  9.87 &  \small  0.00 &  \small  0.00 &  \small 75.25 &  \small 24.75 &  \small  0.00 &  \small  0.00 &  \small 69.98 &  \small 30.02 &  \small  0.00 &  \small  0.00
          \\
 \small345 &  \small360 &  \small 90.43 &  \small  9.57 &  \small  0.00 &  \small  0.00 &  \small 76.01 &  \small 23.99 &  \small  0.00 &  \small  0.00 &  \small 71.03 &  \small 28.97 &  \small  0.00 &  \small  0.00
          \\
 \small360 &  \small375 &  \small 90.69 &  \small  9.31 &  \small  0.00 &  \small  0.00 &  \small 76.73 &  \small 23.27 &  \small  0.00 &  \small  0.00 &  \small 72.00 &  \small 28.00 &  \small  0.00 &  \small  0.00
          \\
 \small375 &  \small390 &  \small 90.95 &  \small  9.05 &  \small  0.00 &  \small  0.00 &  \small 77.37 &  \small 22.63 &  \small  0.00 &  \small  0.00 &  \small 72.88 &  \small 27.12 &  \small  0.00 &  \small  0.00
          \\
 \small390 &  \small405 &  \small 91.19 &  \small  8.81 &  \small  0.00 &  \small  0.00 &  \small 77.97 &  \small 22.03 &  \small  0.00 &  \small  0.00 &  \small 73.70 &  \small 26.30 &  \small  0.00 &  \small  0.00
          \\
 \small405 &  \small420 &  \small 91.40 &  \small  8.60 &  \small  0.00 &  \small  0.00 &  \small 78.52 &  \small 21.48 &  \small  0.00 &  \small  0.00 &  \small 74.44 &  \small 25.56 &  \small  0.00 &  \small  0.00
          \\
 \small420 &  \small435 &  \small 91.60 &  \small  8.40 &  \small  0.00 &  \small  0.00 &  \small 79.06 &  \small 20.94 &  \small  0.00 &  \small  0.00 &  \small 75.16 &  \small 24.84 &  \small  0.00 &  \small  0.00
          \\
 \small435 &  \small450 &  \small 91.80 &  \small  8.20 &  \small  0.00 &  \small  0.00 &  \small 79.56 &  \small 20.44 &  \small  0.00 &  \small  0.00 &  \small 75.85 &  \small 24.15 &  \small  0.00 &  \small  0.00
          \\
 \small450 &  \small465 &  \small 91.98 &  \small  8.02 &  \small  0.00 &  \small  0.00 &  \small 80.03 &  \small 19.97 &  \small  0.00 &  \small  0.00 &  \small 76.48 &  \small 23.52 &  \small  0.00 &  \small  0.00
          \\
 \small465 &  \small480 &  \small 92.16 &  \small  7.84 &  \small  0.00 &  \small  0.00 &  \small 80.50 &  \small 19.50 &  \small  0.00 &  \small  0.00 &  \small 77.11 &  \small 22.89 &  \small  0.00 &  \small  0.00
          \\
 \small480 &  \small495 &  \small 92.34 &  \small  7.66 &  \small  0.00 &  \small  0.00 &  \small 80.96 &  \small 19.04 &  \small  0.00 &  \small  0.00 &  \small 77.71 &  \small 22.29 &  \small  0.00 &  \small  0.00
          \\
 \small495 &  \small510 &  \small 92.49 &  \small  7.51 &  \small  0.00 &  \small  0.00 &  \small 81.33 &  \small 18.67 &  \small  0.00 &  \small  0.00 &  \small 78.22 &  \small 21.78 &  \small  0.00 &  \small  0.00
          \\
 \small510 &  \small525 &  \small 92.63 &  \small  7.37 &  \small  0.00 &  \small  0.00 &  \small 81.70 &  \small 18.30 &  \small  0.00 &  \small  0.00 &  \small 78.71 &  \small 21.29 &  \small  0.00 &  \small  0.00
          \\
 \small525 &  \small540 &  \small 92.77 &  \small  7.23 &  \small  0.00 &  \small  0.00 &  \small 82.05 &  \small 17.95 &  \small  0.00 &  \small  0.00 &  \small 79.19 &  \small 20.81 &  \small  0.00 &  \small  0.00
          \\
 \small540 &  \small555 &  \small 92.89 &  \small  7.11 &  \small  0.00 &  \small  0.00 &  \small 82.37 &  \small 17.63 &  \small  0.00 &  \small  0.00 &  \small 79.62 &  \small 20.38 &  \small  0.00 &  \small  0.00
          \\
 \small555 &  \small570 &  \small 93.01 &  \small  6.99 &  \small  0.00 &  \small  0.00 &  \small 82.68 &  \small 17.32 &  \small  0.00 &  \small  0.00 &  \small 80.03 &  \small 19.97 &  \small  0.00 &  \small  0.00
          \\
 \small570 &  \small585 &  \small 93.12 &  \small  6.88 &  \small  0.00 &  \small  0.00 &  \small 83.01 &  \small 16.99 &  \small  0.00 &  \small  0.00 &  \small 80.46 &  \small 19.54 &  \small  0.00 &  \small  0.00
          \\
 \small585 &  \small600 &  \small 93.24 &  \small  6.76 &  \small  0.00 &  \small  0.00 &  \small 83.31 &  \small 16.69 &  \small  0.00 &  \small  0.00 &  \small 80.87 &  \small 19.13 &  \small  0.00 &  \small  0.00
          \\
 \small600 &  \small615 &  \small 93.35 &  \small  6.65 &  \small  0.00 &  \small  0.00 &  \small 83.61 &  \small 16.39 &  \small  0.00 &  \small  0.00 &  \small 81.27 &  \small 18.73 &  \small  0.00 &  \small  0.00
          \\
 \small615 &  \small630 &  \small 93.47 &  \small  6.53 &  \small  0.00 &  \small  0.00 &  \small 83.90 &  \small 16.10 &  \small  0.00 &  \small  0.00 &  \small 81.63 &  \small 18.37 &  \small  0.00 &  \small  0.00
          \\
 \small630 &  \small645 &  \small 93.57 &  \small  6.43 &  \small  0.00 &  \small  0.00 &  \small 84.17 &  \small 15.83 &  \small  0.00 &  \small  0.00 &  \small 81.97 &  \small 18.03 &  \small  0.00 &  \small  0.00
          \\
 \small645 &  \small660 &  \small 93.67 &  \small  6.33 &  \small  0.00 &  \small  0.00 &  \small 84.44 &  \small 15.56 &  \small  0.00 &  \small  0.00 &  \small 82.32 &  \small 17.68 &  \small  0.00 &  \small  0.00
          \\
 \small660 &  \small675 &  \small 93.77 &  \small  6.23 &  \small  0.00 &  \small  0.00 &  \small 84.71 &  \small 15.29 &  \small  0.00 &  \small  0.00 &  \small 82.66 &  \small 17.34 &  \small  0.00 &  \small  0.00
          \\
 \small675 &  \small690 &  \small 93.86 &  \small  6.14 &  \small  0.00 &  \small  0.00 &  \small 84.94 &  \small 15.06 &  \small  0.00 &  \small  0.00 &  \small 82.97 &  \small 17.03 &  \small  0.00 &  \small  0.00
          \\
 \small690 &  \small705 &  \small 93.96 &  \small  6.04 &  \small  0.00 &  \small  0.00 &  \small 85.17 &  \small 14.83 &  \small  0.00 &  \small  0.00 &  \small 83.28 &  \small 16.72 &  \small  0.00 &  \small  0.00
          \\
 \small705 &  \small720 &  \small 94.05 &  \small  5.95 &  \small  0.00 &  \small  0.00 &  \small 85.37 &  \small 14.63 &  \small  0.00 &  \small  0.00 &  \small 83.54 &  \small 16.46 &  \small  0.00 &  \small  0.00
          \\
 \small720 &  \small735 &  \small 94.12 &  \small  5.88 &  \small  0.00 &  \small  0.00 &  \small 85.56 &  \small 14.44 &  \small  0.00 &  \small  0.00 &  \small 83.78 &  \small 16.22 &  \small  0.00 &  \small  0.00
          \\
\label{tabulatedresults1}
\end{tabular}
\end{table*}

\begin{table*}
\centering
\caption{~Tabulated results from the 8-hour continuous observation case (Fig. \ref{eighthourplot}), the 16-hour continuous observation case (Fig. \ref{sixteenhourplot}) and the difference between F5-M5 spectral types (Fig. \ref{f5m5diffs}), using a bin size of 15 days and taking the median of all results within that bin.  Columns 1 and 2 are the bin range.  P(0), P(1), P(2) and P($\ge$3) are the probabilities of observing zero transits, a single-transit, a double-transit or a multi-transit for planets with those periods, given in percent ($\%$).  A negative percentage means the probability in the M5 spectral-type case was larger than in the F5 spectral-type case.  Table is available in electronic format upon request.}
\begin{tabular}{@{}l l l l l l l l l l l l l l}
\hline
\small{$P_{min}$} & \small{$P_{max}$} & \small{8~H} & & & & \small{16~H} & & & & \small{F5-M5} & & &
\\
   \small{(days)}     &    \small{(days)}     & \small{P(0)} & \small{P(1)} & \small{P(2)} & \small{P($\ge$3)} & \small{P(0)} & \small{P(1)} & \small{P(2)} & \small{P($\ge$3)} & \small{P(0)} & \small{P(1)} & \small{P(2)} & \small{P($\ge$3)}
\\
\hline
 \small  0 &  \small 15 &  \small  1.01 &  \small  0.06 &  \small  0.06 &  \small 98.87 &  \small  0.08 &  \small  0.01 &  \small  0.01 &  \small 99.90 &  \small-36.24 &  \small-11.94 &  \small -9.61 &  \small 57.79 \\
 \small 15 &  \small 30 &  \small  0.80 &  \small  0.25 &  \small  0.30 &  \small 98.66 &  \small  0.09 &  \small  0.01 &  \small  0.00 &  \small 99.89 &  \small-10.34 &  \small-20.53 &  \small -2.96 &  \small 33.83 \\
 \small 30 &  \small 45 &  \small  2.38 &  \small  0.72 &  \small  0.79 &  \small 96.10 &  \small  0.04 &  \small  0.06 &  \small  0.08 &  \small 99.81 &  \small-18.14 &  \small-12.71 &  \small 16.37 &  \small 14.48 \\
 \small 45 &  \small 60 &  \small  1.75 &  \small  1.33 &  \small  3.86 &  \small 93.06 &  \small  0.01 &  \small  0.03 &  \small  0.03 &  \small 99.93 &  \small-22.17 &  \small -0.94 &  \small 17.51 &  \small  5.61 \\
 \small 60 &  \small 75 &  \small  2.03 &  \small  1.38 &  \small  9.63 &  \small 86.96 &  \small  0.08 &  \small  0.00 &  \small  0.00 &  \small 99.92 &  \small-23.29 &  \small  6.17 &  \small 15.83 &  \small  1.28 \\
 \small 75 &  \small 90 &  \small  2.49 &  \small  3.70 &  \small 24.58 &  \small 69.22 &  \small  0.00 &  \small  0.00 &  \small  0.00 &  \small100.00 &  \small-23.73 &  \small 11.94 &  \small 11.77 &  \small  0.03 \\
 \small 90 &  \small105 &  \small  4.21 &  \small  5.83 &  \small 36.17 &  \small 53.79 &  \small  0.00 &  \small  0.00 &  \small  0.00 &  \small100.00 &  \small-23.62 &  \small 15.90 &  \small  7.72 &  \small  0.00 \\
 \small105 &  \small120 &  \small  3.78 &  \small 10.74 &  \small 44.10 &  \small 41.38 &  \small  0.00 &  \small  0.00 &  \small  0.00 &  \small100.00 &  \small-23.53 &  \small 19.23 &  \small  4.30 &  \small  0.00 \\
 \small120 &  \small135 &  \small  3.69 &  \small 16.21 &  \small 50.00 &  \small 30.10 &  \small  0.00 &  \small  0.00 &  \small 13.52 &  \small 86.48 &  \small-23.20 &  \small 21.16 &  \small  2.04 &  \small  0.00 \\
 \small135 &  \small150 &  \small  3.94 &  \small 25.65 &  \small 47.70 &  \small 22.71 &  \small  0.00 &  \small  0.00 &  \small 43.53 &  \small 56.47 &  \small-22.66 &  \small 22.16 &  \small  0.50 &  \small  0.00 \\
 \small150 &  \small165 &  \small  5.49 &  \small 27.42 &  \small 54.05 &  \small 13.04 &  \small  0.00 &  \small  0.00 &  \small 68.26 &  \small 31.74 &  \small-21.32 &  \small 21.25 &  \small  0.07 &  \small  0.00 \\
 \small165 &  \small180 &  \small  5.16 &  \small 32.26 &  \small 57.53 &  \small  5.04 &  \small  0.00 &  \small  0.00 &  \small 88.57 &  \small 11.43 &  \small-20.06 &  \small 20.06 &  \small  0.00 &  \small  0.00 \\
 \small180 &  \small195 &  \small  4.59 &  \small 38.67 &  \small 56.66 &  \small  0.08 &  \small  0.00 &  \small  5.63 &  \small 94.19 &  \small  0.18 &  \small-18.96 &  \small 18.96 &  \small  0.00 &  \small  0.00 \\
 \small195 &  \small210 &  \small  9.20 &  \small 38.31 &  \small 52.49 &  \small  0.00 &  \small  0.00 &  \small 19.89 &  \small 80.11 &  \small  0.00 &  \small-17.99 &  \small 17.99 &  \small  0.00 &  \small  0.00 \\
 \small210 &  \small225 &  \small  7.05 &  \small 50.22 &  \small 42.74 &  \small  0.00 &  \small  0.00 &  \small 32.12 &  \small 67.88 &  \small  0.00 &  \small-16.90 &  \small 16.90 &  \small  0.00 &  \small  0.00 \\
 \small225 &  \small240 &  \small 10.39 &  \small 50.35 &  \small 39.26 &  \small  0.00 &  \small  0.00 &  \small 42.77 &  \small 57.23 &  \small  0.00 &  \small-16.12 &  \small 16.12 &  \small  0.00 &  \small  0.00 \\
 \small240 &  \small255 &  \small 10.35 &  \small 56.53 &  \small 33.12 &  \small  0.00 &  \small  0.00 &  \small 52.39 &  \small 47.61 &  \small  0.00 &  \small-15.23 &  \small 15.23 &  \small  0.00 &  \small  0.00 \\
 \small255 &  \small270 &  \small 10.40 &  \small 62.08 &  \small 27.52 &  \small  0.00 &  \small  0.00 &  \small 61.09 &  \small 38.91 &  \small  0.00 &  \small-14.56 &  \small 14.56 &  \small  0.00 &  \small  0.00 \\
 \small270 &  \small285 &  \small 10.49 &  \small 66.89 &  \small 22.62 &  \small  0.00 &  \small  0.00 &  \small 68.68 &  \small 31.32 &  \small  0.00 &  \small-13.91 &  \small 13.91 &  \small  0.00 &  \small  0.00 \\
 \small285 &  \small300 &  \small 10.91 &  \small 70.45 &  \small 18.64 &  \small  0.00 &  \small  0.00 &  \small 75.28 &  \small 24.72 &  \small  0.00 &  \small-13.28 &  \small 13.28 &  \small  0.00 &  \small  0.00 \\
 \small300 &  \small315 &  \small 10.69 &  \small 74.97 &  \small 14.34 &  \small  0.00 &  \small  0.00 &  \small 81.28 &  \small 18.72 &  \small  0.00 &  \small-12.74 &  \small 12.74 &  \small  0.00 &  \small  0.00 \\
 \small315 &  \small330 &  \small 10.37 &  \small 79.28 &  \small 10.34 &  \small  0.00 &  \small  0.00 &  \small 86.73 &  \small 13.27 &  \small  0.00 &  \small-12.24 &  \small 12.24 &  \small  0.00 &  \small  0.00 \\
 \small330 &  \small345 &  \small 10.28 &  \small 82.83 &  \small  6.89 &  \small  0.00 &  \small  0.00 &  \small 91.67 &  \small  8.33 &  \small  0.00 &  \small-11.78 &  \small 11.78 &  \small  0.00 &  \small  0.00 \\
 \small345 &  \small360 &  \small  9.58 &  \small 87.50 &  \small  2.92 &  \small  0.00 &  \small  0.00 &  \small 96.40 &  \small  3.60 &  \small  0.00 &  \small-11.36 &  \small 11.36 &  \small  0.00 &  \small  0.00 \\
 \small360 &  \small375 &  \small  9.75 &  \small 90.07 &  \small  0.18 &  \small  0.00 &  \small  0.79 &  \small 99.00 &  \small  0.21 &  \small  0.00 &  \small-10.92 &  \small 10.92 &  \small  0.00 &  \small  0.00 \\
 \small375 &  \small390 &  \small 12.22 &  \small 87.78 &  \small  0.00 &  \small  0.00 &  \small  4.43 &  \small 95.57 &  \small  0.00 &  \small  0.00 &  \small-10.57 &  \small 10.57 &  \small  0.00 &  \small  0.00 \\
 \small390 &  \small405 &  \small 14.74 &  \small 85.26 &  \small  0.00 &  \small  0.00 &  \small  7.98 &  \small 92.02 &  \small  0.00 &  \small  0.00 &  \small-10.23 &  \small 10.23 &  \small  0.00 &  \small  0.00 \\
 \small405 &  \small420 &  \small 17.02 &  \small 82.98 &  \small  0.00 &  \small  0.00 &  \small 11.19 &  \small 88.81 &  \small  0.00 &  \small  0.00 &  \small -9.92 &  \small  9.92 &  \small  0.00 &  \small  0.00 \\
 \small420 &  \small435 &  \small 19.25 &  \small 80.75 &  \small  0.00 &  \small  0.00 &  \small 14.29 &  \small 85.71 &  \small  0.00 &  \small  0.00 &  \small -9.61 &  \small  9.61 &  \small  0.00 &  \small  0.00 \\
 \small435 &  \small450 &  \small 21.38 &  \small 78.62 &  \small  0.00 &  \small  0.00 &  \small 17.28 &  \small 82.72 &  \small  0.00 &  \small  0.00 &  \small -9.33 &  \small  9.33 &  \small  0.00 &  \small  0.00 \\
 \small450 &  \small465 &  \small 23.33 &  \small 76.67 &  \small  0.00 &  \small  0.00 &  \small 19.99 &  \small 80.01 &  \small  0.00 &  \small  0.00 &  \small -9.07 &  \small  9.07 &  \small  0.00 &  \small  0.00 \\
 \small465 &  \small480 &  \small 24.36 &  \small 75.64 &  \small  0.00 &  \small  0.00 &  \small 22.60 &  \small 77.40 &  \small  0.00 &  \small  0.00 &  \small -8.84 &  \small  8.84 &  \small  0.00 &  \small  0.00 \\
 \small480 &  \small495 &  \small 25.13 &  \small 74.87 &  \small  0.00 &  \small  0.00 &  \small 25.13 &  \small 74.87 &  \small  0.00 &  \small  0.00 &  \small -8.59 &  \small  8.59 &  \small  0.00 &  \small  0.00 \\
 \small495 &  \small510 &  \small 27.26 &  \small 72.74 &  \small  0.00 &  \small  0.00 &  \small 27.26 &  \small 72.74 &  \small  0.00 &  \small  0.00 &  \small -8.36 &  \small  8.36 &  \small  0.00 &  \small  0.00 \\
 \small510 &  \small525 &  \small 29.32 &  \small 70.68 &  \small  0.00 &  \small  0.00 &  \small 29.32 &  \small 70.68 &  \small  0.00 &  \small  0.00 &  \small -8.17 &  \small  8.17 &  \small  0.00 &  \small  0.00 \\
 \small525 &  \small540 &  \small 31.32 &  \small 68.68 &  \small  0.00 &  \small  0.00 &  \small 31.32 &  \small 68.68 &  \small  0.00 &  \small  0.00 &  \small -7.96 &  \small  7.96 &  \small  0.00 &  \small  0.00 \\
 \small540 &  \small555 &  \small 33.12 &  \small 66.88 &  \small  0.00 &  \small  0.00 &  \small 33.12 &  \small 66.88 &  \small  0.00 &  \small  0.00 &  \small -7.77 &  \small  7.77 &  \small  0.00 &  \small  0.00 \\
 \small555 &  \small570 &  \small 34.87 &  \small 65.13 &  \small  0.00 &  \small  0.00 &  \small 34.87 &  \small 65.13 &  \small  0.00 &  \small  0.00 &  \small -7.60 &  \small  7.60 &  \small  0.00 &  \small  0.00 \\
 \small570 &  \small585 &  \small 36.57 &  \small 63.43 &  \small  0.00 &  \small  0.00 &  \small 36.57 &  \small 63.43 &  \small  0.00 &  \small  0.00 &  \small -7.42 &  \small  7.42 &  \small  0.00 &  \small  0.00 \\
 \small585 &  \small600 &  \small 38.23 &  \small 61.77 &  \small  0.00 &  \small  0.00 &  \small 38.23 &  \small 61.77 &  \small  0.00 &  \small  0.00 &  \small -7.25 &  \small  7.25 &  \small  0.00 &  \small  0.00 \\
 \small600 &  \small615 &  \small 39.84 &  \small 60.16 &  \small  0.00 &  \small  0.00 &  \small 39.84 &  \small 60.16 &  \small  0.00 &  \small  0.00 &  \small -7.08 &  \small  7.08 &  \small  0.00 &  \small  0.00 \\
 \small615 &  \small630 &  \small 41.29 &  \small 58.71 &  \small  0.00 &  \small  0.00 &  \small 41.29 &  \small 58.71 &  \small  0.00 &  \small  0.00 &  \small -6.92 &  \small  6.92 &  \small  0.00 &  \small  0.00 \\
 \small630 &  \small645 &  \small 42.57 &  \small 57.43 &  \small  0.00 &  \small  0.00 &  \small 42.57 &  \small 57.43 &  \small  0.00 &  \small  0.00 &  \small -6.78 &  \small  6.78 &  \small  0.00 &  \small  0.00 \\
 \small645 &  \small660 &  \small 43.94 &  \small 56.06 &  \small  0.00 &  \small  0.00 &  \small 43.94 &  \small 56.06 &  \small  0.00 &  \small  0.00 &  \small -6.62 &  \small  6.62 &  \small  0.00 &  \small  0.00 \\
 \small660 &  \small675 &  \small 45.28 &  \small 54.72 &  \small  0.00 &  \small  0.00 &  \small 45.28 &  \small 54.72 &  \small  0.00 &  \small  0.00 &  \small -6.49 &  \small  6.49 &  \small  0.00 &  \small  0.00 \\
 \small675 &  \small690 &  \small 46.48 &  \small 53.52 &  \small  0.00 &  \small  0.00 &  \small 46.48 &  \small 53.52 &  \small  0.00 &  \small  0.00 &  \small -6.35 &  \small  6.35 &  \small  0.00 &  \small  0.00 \\
 \small690 &  \small705 &  \small 47.64 &  \small 52.36 &  \small  0.00 &  \small  0.00 &  \small 47.64 &  \small 52.36 &  \small  0.00 &  \small  0.00 &  \small -6.20 &  \small  6.20 &  \small  0.00 &  \small  0.00 \\
 \small705 &  \small720 &  \small 48.67 &  \small 51.33 &  \small  0.00 &  \small  0.00 &  \small 48.67 &  \small 51.33 &  \small  0.00 &  \small  0.00 &  \small -6.10 &  \small  6.10 &  \small  0.00 &  \small  0.00 \\
 \small720 &  \small735 &  \small 49.57 &  \small 50.43 &  \small  0.00 &  \small  0.00 &  \small 49.57 &  \small 50.43 &  \small  0.00 &  \small  0.00 &  \small -6.00 &  \small  6.00 &  \small  0.00 &  \small  0.00 \\

\label{tabulatedresults2}
\end{tabular}
\end{table*}


\label{lastpage}


\begin{thebibliography}{}

\bibitem[\protect\citeauthoryear{Allen \& Cox}{2000}]{all2000} Arthur N. \& Cox E., Allen's Astrophysical Quantities, 4th ed. (AIP/Springer-Verlag, New York, 2001), p.389

\bibitem[\protect\citeauthoryear{Alonso et al.}{2004}]{alo2004} Alonso R., et al., 2004, ApJ, 613L, 153A

\bibitem[\protect\citeauthoryear{Bakos et al.}{2004}]{bak2004} Bakos G., Noyes R.W., Kov\'{a}cs G., Stanek K.Z., Sasselov D.D., Domsa I., 2004, PASP, 116, 266

\bibitem[\protect\citeauthoryear{Bakos et al.}{2007a}]{bak2007a} Bakos G.\'{A}., et al., 2007a, ApJ, 656, 552B

\bibitem[\protect\citeauthoryear{Bakos et al.}{2007b}]{bak2007b} Bakos G.\'{A}., et al., 2007b, ApJ, 670, 826B

\bibitem[\protect\citeauthoryear{Barbieri et al.}{2007}]{barb2007} Barbieri M., et al., 2007, A \& A, 476L, 13B

\bibitem[\protect\citeauthoryear{Barnes}{2007}]{bar2007} Barnes J.W., 2007, PASP, 119, 986B

\bibitem[\protect\citeauthoryear{Barnes \& O'Brien}{2002}]{bar2002} Barnes J.W. \& O'Brien D.P., 2002, ApJ, 575, 1087

\bibitem[\protect\citeauthoryear{Bayless \& Orosz}{2006}] {bay2006} Bayless A.J. \& Orosz J.A., 2006, ApJ, 651, 1155

\bibitem[\protect\citeauthoryear{Borucki \& Summers}{1984}]{bor1984} Borucki W.J. \& Summers A.L., 1984, Icarus, 58, 121

\bibitem[\protect\citeauthoryear{Borucki et al.}{2001}]{bor2001} Borucki W., Caldwell D., Koch D., Webster L., Jenkins J., Ninkov Z., Showen R., 2001, PASP, 113, 439

\bibitem[\protect\citeauthoryear{Bouchy et al.}{2004}]{bou2004} Bouchy F., Pont F., Santos N.C., Melo C., Mayor M., Queloz D., Udry S., 2004, A\&A, 421L, 13

\bibitem[\protect\citeauthoryear{Butler et al.}{2006}]{but2006} Butler R.P., et al., 2006, ApJ, 646, 505B

\bibitem[\protect\citeauthoryear{Brown}{2003}]{bro2003} Brown T.M., 2003, ApJ, 593L, 125

\bibitem[\protect\citeauthoryear{Brown \& Charbonneau}{1999}]{bro1999} Brown T.M. \& Charbonneau D., 1999, BAAS, 31, 1534

\bibitem[\protect\citeauthoryear{Cameron et al.}{2007}]{cam2007} Cameron A. Collier, et al., 2007, MNRAS, 375, 951C

\bibitem[\protect\citeauthoryear{Charbonneau et al.}{2002}]{cha2002} Charbonneau D., Brown T.M., Noyes R.W., Gilliland R.L., 2002, ApJ, 568, 377

\bibitem[\protect\citeauthoryear{Cox}{2000a}]{cox2000table} Cox A.N., 2000a, in Allen's Astrophysical Quantities (New York: AIP), 389

\bibitem[\protect\citeauthoryear{Cox}{2000b}]{cox2000} Cox A.N., 2000b, in Allen's Astrophysical Quantities (New York: AIP), 355-357

\bibitem[\protect\citeauthoryear{Doyle \& Deeg}{2002}]{doy2002} Doyle L.R. \& Deeg H.-J., 2002, arXiv:astro-ph/0306087 [astro-ph]

\bibitem[\protect\citeauthoryear{Duquennoy \& Mayor}{1991}]{duq1991} Duquennoy A. \& Mayor M., 1991, A\&A, 248, 485

\bibitem[\protect\citeauthoryear{Eggenberger et al.}{2004}]{egg2004} Eggenberger A., Halbwachs J.-L., Udry S., Mayor M., 2004, RevMexAA, 21, 28

\bibitem[\protect\citeauthoryear{Fischer et al.}{2007}]{fis2007} Fischer D.A., et al., arXiv:0704.1191v2 [astro-ph], 2007

\bibitem[\protect\citeauthoryear{Gould et al.}{2003a}]{goufis2003} Gould A., Ford E.B., Fischer D.A., 2003a, ApJ, 591L, 155

\bibitem[\protect\citeauthoryear{Gould et al.}{2003b}]{goupep2003} Gould A., Pepper J., DePoy D.L., 2003b, ApJ, 594, 533

\bibitem[\protect\citeauthoryear{Halbwachs et al.}{2003}]{hal2003} Halbwachs J.L., Mayor M., Udry S., Arenou F., 2003, A\&A, 397, 159

\bibitem[\protect\citeauthoryear{Jones et al.}{2006}]{jon2006} Jones B.W., Sleep P.N., Underwood D.R., 2006, ApJ, 649, 1010

\bibitem[\protect\citeauthoryear{Kane et al.}{2007}]{kan2007} Kane S.R., et al., 2007, MNRAS, arXiv:0711.2581 [astro-ph]

\bibitem[\protect\citeauthoryear{Kasting et al.}{1993}]{kas1993} Kasting J.F., Whitmire D.P., Reynolds R.T., 1993, Icarus, 101, 108

\bibitem[\protect\citeauthoryear{Konacki et al.}{2004}]{kon2004} Konacki M., et al., 2004, ApJ, 609L, 37

\bibitem[\protect\citeauthoryear{Kov\'{a}cs et al.}{2002}]{kov2002} Kov\'{a}cs G., Zucker S., Mazeh T., 2002, A\&A, 391, 369

\bibitem[\protect\citeauthoryear{Lastennet \& Valls-Gabaud}{2002}]{las2002} Lastennet E. \& Valls-Gabaud D., 2002, A\&A, 396, 551

\bibitem[\protect\citeauthoryear{Liang et al.}{2003}]{lia2003} Liang M., Parkinson C.D., Lee A.Y.-T., Yung Y.L., Seager S., 2003, ApJ, 596, L247

\bibitem[\protect\citeauthoryear{L\'{o}pez-Morales \& Clemens}{2004}]{lop2004} L\'{o}pez-Morales M. \& Clemens J.C., 2004, PASP, 116, 22

\bibitem[\protect\citeauthoryear{McCullough \& Burke}{2007}]{mcc2007} McCullough P.R. \& Burke C.J., 2007, ASPC, 366, 70M

\bibitem[\protect\citeauthoryear{McCullough et al.}{2005}]{mcc2005} McCullough P.R., Stys J.E., Valenti J.A., Fleming S.W., Janes K.A., Heasley J.N., 2005, PASP, 117, 783

\bibitem[\protect\citeauthoryear{McCullough et al.}{2006}]{mcc2006} McCullough P.R., et al., ApJ, 2006, 648, 1228M

\bibitem[\protect\citeauthoryear{Moutou \& Pont}{2006}]{mou2006} Moutou C. \& Pont F., Formation Plan\'{e}taire et Exoplan\`{e}tes, Ecole CNRS de Goutelas XXVIII (2005), 2006, Volume 28, pg. 55

\bibitem[\protect\citeauthoryear{Nutzman \& Charbonneau}{2007}]{nut2007} Nutzman P. \& Charbonneau D., arXiv:0709.2879v1 [astro-ph], 2007

\bibitem[\protect\citeauthoryear{Pollacco et al.}{2006}]{pol2006} Pollacco D.L., et al., 2006, PASP, 118, 1407

\bibitem[\protect\citeauthoryear{Pont et al.}{2006}]{pon2006} Pont F., Zucker S., Queloz D., 2006, MNRAS, 373, 231

\bibitem[\protect\citeauthoryear{Robinson et al.}{2007}]{rob2007} Robinson S.E., et al., 2007, ApJ, 670, 1391

\bibitem[\protect\citeauthoryear{Sartoretti \& Schneider}{1999}]{sar1999} Sartoretti P. \& Schneider J., 1999, Astron. Astrophys. Suppl. Ser., 134, 553

\bibitem[\protect\citeauthoryear{Seager \& Mall\'{e}n-Ornelas}{2003}]{seag2003} Seager S. \& Mall\'{e}n-Ornelas G., 2003, ApJ, 585, 1038

\bibitem[\protect\citeauthoryear{Seagroves et al.}{2003}]{sea2003} Seagroves S., Harker J., Laughlin G., Lacy J., Castellano T., 2003, PASP, 115, 1355

\bibitem[\protect\citeauthoryear{Stassun et al.}{2006}]{sta2006} Stassun K.G., Mathieu R.D., Valenti J.A., 2006, Nature, 440, 311S

\bibitem[\protect\citeauthoryear{Turnbull \& Tarter}{2003}]{tur2003} Turnbull M.C. \& Tarter J.C., 2003, ApJS, 149, 423

\bibitem[\protect\citeauthoryear{Torres et al.}{2004}]{tor2004} Torres G., Konacki M., Sasselov D.D., Jha S., 2004, ApJ, 609, 1071

\bibitem[\protect\citeauthoryear{Udalski et al.}{2002}]{uda2002} Udalski A., \.{Z}ebru\'{n} K., Szyma\'{n}ski M., Kubiak M., Soszy\'{n}ski I., Szewczyk O., Wyrzykowski \L., Pietrzy\'{n}ski G., 2002, Acta. Astron., 52, 115

\bibitem[\protect\citeauthoryear{Udry \& Santos}{2007}]{udr2007} Udry S. \& Santos N. C., 2007, ARA\&A, 45, 397U

\bibitem[\protect\citeauthoryear{Udry et al.}{2003}]{udr2003} Udry S., Mayor M., Santos N.C., 2003, A\&A, 407, 369

\bibitem[\protect\citeauthoryear{Vidal-Madjar et al.}{2003}]{vid2003} Vidal-Madjar A., Lecavelier des Etangs A., D\'{e}sert J.-M., Ballester G.E., Ferlet R., H\'{e}brard G., Mayor M., 2003, Nature, 422, 143

\bibitem[\protect\citeauthoryear{Vidal-Madjar et al.}{2004}]{vid2004} Vidal-Madjar A., et al., 2004, ApJ, 604, L69

\bibitem[\protect\citeauthoryear{Wyithe \& Wilson}{2001}]{wyi2001} Wyithe J.S.B. \& Wilson R.E., 2001, ApJ, 559, 260

\end{thebibliography}
\end{document}